\documentclass[sigconf, nonacm=true]{acmart}

\copyrightyear{2025}
\acmYear{2025}
\setcopyright{rightsretained}
\acmConference[WSDM '25]{Proceedings of the Eighteenth ACM International Conference on Web Search and Data Mining}{March 10--14, 2025}{Hannover, Germany}
\acmBooktitle{Proceedings of the Eighteenth ACM International Conference on Web Search and Data Mining (WSDM '25), March 10--14, 2025, Hannover, Germany}
\acmDOI{10.1145/3701551.3703491}
\acmISBN{979-8-4007-1329-3/25/03}

\makeatletter
\gdef\@copyrightpermission{
  \begin{minipage}{0.3\columnwidth}
   \href{https://creativecommons.org/licenses/by-nc-sa/4.0/}{\includegraphics[width=0.90\textwidth]{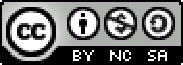}}
  \end{minipage}\hfill
  \begin{minipage}{0.7\columnwidth}
   \href{https://creativecommons.org/licenses/by-nc-sa/4.0/}{This work is licensed under a Creative Commons Attribution-NonCommercial-ShareAlike International 4.0 License.}
  \end{minipage}
  \vspace{5pt}
}
\makeatother

\usepackage{makecell}
\usepackage{float}
\usepackage{multirow}
\usepackage{subcaption}
\usepackage{amsthm}
\usepackage{hyperref}
\usepackage[capitalise]{cleveref}
\usepackage{enumitem}
\usepackage{balance}

\usepackage{algorithm} 
\usepackage{algpseudocode} 

\theoremstyle{plain}
\newtheorem{theorem}{Theorem}[section]
\newtheorem{lemma}[theorem]{Lemma}
\newtheorem{corollary}[theorem]{Corollary}

\theoremstyle{definition}
\newtheorem{definition}[theorem]{Definition}

\theoremstyle{remark}

\crefname{theorem}{Theorem}{Theorems}
\crefname{definition}{Definition}{Definitions}

\newcommand{\ie}{i.e.,}
\newcommand{\eg}{e.g.,}
\newcommand{\wrt}{w.r.t.}
\newcommand{\Wlog}{w.l.o.g.,}

\newcommand{\Whp}{\emph{w.h.p.\ }}

\newcommand{\minh}[1]{\mbox{\sc{MinHash}} \left( #1 \right)}
\newcommand{\minht}{\mbox{\sc{MinHash}}}
\newcommand{\kminht}{\mbox{$k$-\sc{MinHash}}}
\newcommand{\Jsim}{\mbox{\sc{J}}}
\newcommand{\Smallest}{\mbox{\sc{Smallest}}}
\newcommand{\vanilla}{\emph{Vanilla-\minht}}

\newcommand{\sH}{\mathcal H}

\newcommand{\buff}{\mathcal{B}}
\newcommand{\struct}{\mathcal{S}}

\newcommand{\textbello}[1]{\textsc{#1}}

\newcommand{\fullversion}{1}
\newcommand{\iffull}[2]{%
  \if\fullversion1 #1\else#2\fi
}

\begin{document}

\title{Maintaining \texorpdfstring{$k$}{k}-MinHash Signatures over Fully-Dynamic Data Streams with Recovery}

\author{Andrea Clementi}
\authornote{All the authors contributed equally to this research.}
\email{clementi@mat.uniroma2.it}
\orcid{0000-0002-9521-2457}
\affiliation{%
  \institution{University of Rome ``Tor Vergata''}
  \department{Department of Enterprise Engineering}
  \city{Rome}
  \country{Italy}
}

\author{Luciano Gualà}
\authornotemark[1]
\email{guala@mat.uniroma2.it}
\orcid{0000-0001-6976-5579}
\affiliation{%
  \institution{University of Rome ``Tor Vergata''}
  \department{Department of Enterprise Engineering}
  \city{Rome}
  \country{Italy}
}

\author{Luca Pepè Sciarria}
\authornotemark[1]
\authornote{This work has been supported by the Spoke 1 ``FutureHPC \& BigData'' of the Italian Research Center on High Performance Computing, Big Data and Quantum Computing (ICSC) funded by MUR Missione 4 Componente 2 Investimento 1.4: Potenziamento strutture di ricerca e creazione di ``campioni nazionali'' di R\& S (M4C2-19) - Next Generation EU (NGEU).}
\email{luca.pepesciarria@gmail.com}
\orcid{0000-0003-4432-6099}
\affiliation{%
  \institution{University of Rome ``Tor Vergata''}
  \department{Department of Enterprise Engineering}
  \city{Rome}
  \country{Italy}
}

\author{Alessandro Straziota}
\authornotemark[1]
\email{alessandro.straziota@uniroma2.it}
\orcid{0009-0008-4543-786X}
\affiliation{%
  \institution{University of Rome ``Tor Vergata''}
  \department{Department of Enterprise Engineering}
  \city{Rome}
  \country{Italy}
}

\renewcommand{\shortauthors}{Andrea Clementi, Luciano Gualà, Luca Pepè Sciarria, \& Alessandro Straziota}

\ccsdesc[500]{Theory of computation~Sketching and sampling}
\ccsdesc[500]{Theory of computation~Data structures design and analysis}
\ccsdesc[500]{Information systems~Data stream mining}

\keywords{Data Sketches, Jaccard Similarity Estimation, Dynamic Data Streams, MinHashing, Probabilistic/Amortized Analysis of Algorithms}

\begin{abstract}
We consider the task of performing Jaccard similarity queries over a large collection of items that are dynamically updated according to a streaming input model. An item here is a subset of a large universe $U$ of elements.
A well-studied approach to address this important problem in data mining is to design \textit{fast-similarity data sketches}. In this paper, we focus on \textit{global solutions} for this problem, \ie\ a single data structure which is able to answer both \textit{Similarity Estimation} and \textit{All-Candidate Pairs} queries, while also dynamically managing an arbitrary, online sequence of element insertions and deletions received in input.

We introduce and provide an in-depth analysis of a dynamic, buffered version of the well-known \kminht\ sketch. This buffered version better manages critical update operations thus significantly reducing the number of times the sketch needs to be rebuilt from scratch using expensive recovery queries.
We prove that the \textit{buffered} \kminht\ uses $O(k \log |U|)$ memory words per subset and that its \textit{amortized} update time per insertion/deletion is $O(k \log |U|)$ \textit{with high probability}. Moreover, our data structure can return the \kminht\ signature of any subset in $O(k)$ time, and this signature is exactly the same signature that would be computed from scratch (and thus the quality of the signature is the same as the one guaranteed by the static \kminht).

Analytical and experimental comparisons with the other, state-of-the-art global solutions for this problem given in [Bury et al.,WSDM'18] show that the \textit{buffered} \kminht\ turns out to be competitive in a wide and relevant range of the online input parameters.
\end{abstract}

\maketitle

\section{Introduction} \label{sec:intro}

A fundamental task in data mining is to detect ``similar'' items in a given large collection \cite{VOS,Leskobook20,MROS}.
Once an effective digital encoding of the items and a suitable notion of item \textit{similarity} have been selected for the considered application (\eg\ items are represented as points of a multidimensional metric space), performing fast similarity queries often requires the use of \textit{compressed} representations of the items also known as \textit{sketches} (a.k.a. \textit{fingerprints} or \textit{signatures}) \cite{blum2020foundations,broder00identifying,Leskobook20,ANF02}. Then, for our purpose, a \textit{sketch} $\mathcal{S}(A)$ of an item $A$ can be informally defined as a data structure, significantly more compact than the \emph{size} of $A$, that somewhat preserves the original similarity between any pair of items \cite{broder00identifying,charikar2002similarity,Leskobook20}. This key property allows us to efficiently perform \emph{Similarity Estimation (SE)} and \textit{All Candidate Pairs} (\textit{ACP}) \cite{dahlgaard2017fast,indyk1998approximate,Leskobook20} over a collection of items: in the former query, the goal is to estimate the similarity between any pair of items given in input, while in the latter query, the goal is to return all \textit{candidate}  item pairs of the collection having similarity at least $\lambda$, where $\lambda >0$ is an input threshold parameter.

Following a large amount of previous works \cite{baeza99modern,ANF02,Leskobook20,manber94shing,VOS}, we consider the general, standard framework in which an item can be represented as a \emph{set} $A$ which is a subset of a fixed, large universe $U$ (\Wlog\ we can assume $U$ be the set $[N] = \{1,2, \ldots, N \}$), and we adopt the \emph{Jaccard similarity} \cite{Leskobook20}:
given $A,B \subseteq U$, the Jaccard similarity $\Jsim(A,B)$ is defined as $\Jsim(A,B) = \vert A \cap B \vert / \vert A \cup B \vert$.

In terms of quality, the goal is to guarantee, at the same time, provably-good similarity estimation, limited memory usage, and fast query time-response even when the set collection is very large. Additionally, data sketches in online systems should efficiently support basic updating operations that arrive according to some \textit{data-streaming} process \cite{BSS20,VOS,MROS,Dynamic_minwise}.

More formally, we study the problem of efficiently maintaining, via suitable sketches, a collection $A_1, \dots, A_m$ of $m$ sets that can be updated according to an input streaming model and for which we want to perform SE and ACP queries in an online fashion. The input stream is a sequence of triples $\langle i,x,o \rangle$, where $i \in [m]$ is the set identifier, $x \in U$ is an element, and $o \in \{-1,+1\}$ specifies whether an insertion or a deletion (denoted as $+1$ and $-1$, respectively) of element $x$ in the set $A_i$ has to be done.

In the above framework, efficient solutions for SE queries on fully-dynamic input streams have been obtained in \cite{MROS,VOS}, while in \cite{BSS20} a sketch scheme has been derived that is also able to answer ACP queries. This latter sketch scheme \cite{BSS20} is particularly relevant to us, since in this work we focus on the design of \textit{global} solutions for Jaccard similarity, namely a single data structure that allows to efficiently answer to \emph{both} SE and ACP queries. For this reason, the solution in \cite{BSS20} will be thoroughly discussed and compared with ours later in \cref{sec:comparison}. However, we point out here that the solution in \cite{BSS20} adopts a classical streaming model in which the sketch algorithm can read the stream of update operations only once while answering queries in an online fashion. Moreover, their solution assumes that the sequence of updates must be \textit{legal}. A data stream is said to be \textit{legal} if an element can be inserted into a set $A$ only if it does not already belong to $A$ and it can be deleted only if it is already inside $A$.

Notice that in real scenarios, non-legal streams typically take place whenever multiple streams coming from different data sources refer to the same set collection. A concrete instance of this scenario arises in web-graph mining where the sets represent the neighborhoods of the vertices and the updates are detected by many autonomous crawlers exploring the same web subgraph \cite{kobayashi2000information,langville2006google}.
To the best of our knowledge, no (global or otherwise) provably-good solution for arbitrary (\ie\ non-legal) streams is available.

\subsection{Our Contribution}

In this paper we adopt a stronger model than the one used in \cite{BSS20} that we call \emph{Data Stream with Recovery}.
We assume that the sets of the collection are stored in a different location (for instance another machine, or the cloud, or a secondary memory) and that the sketch algorithm can ask for the current state of a given set to obtain a new stream of its elements.
Clearly, \emph{recovery queries} come with an \textit{additional} cost of processing such a new stream and their optimization is far from trivial: this is in fact one of the main goals of this paper.

The Data Stream with Recovery model is inspired by real scenarios in which recovery queries are indeed available, like in cloud computing systems \cite{Kathare2022ACS, DynamoDB}, or in the three-layer client server model adopted for network mining in \cite{VOS,MROS,PR00}.
Moreover, we remark that this assumption does not make our problem trivial since we still need to compute and maintain the set sketches in order to perform fast SE and ACP queries.

As a concrete instance of this framework, we mention here the recent work \cite{pang2024similarity} in which the \kminht\ signatures are used to represent the node neighborhoods of a large graph. The \kminht\ is maybe the most popular and effective sketch scheme for Jaccard similarity  \cite{broder1997resemblance,BCFM00} (see \Cref{sc:preliminaries} for its formal definition) which allows both SE and ACP queries.
In \cite{pang2024similarity}, the entire graph is maintained in memory (and thus it can be \textit{queried}) together with the sketches that are used to \textit{speed-up} the similarity computations over many node neighborhoods which in turn are exploited to find dense subgraphs\footnote{We remark that, in \cite{pang2024similarity}, maintaining \kminht\ signatures over dynamic input models is left as an important open problem.}.

Interestingly enough, the Data Stream with Recovery model is intrinsically adopted by other works \cite{BSS20,VOS,MROS} in order to compare the proposed fully-dynamic solutions with a baseline called \vanilla. The latter is a basic fully-dynamic version of the \kminht.
Informally, \vanilla\ explicitly maintains the \kminht\ \emph{signature} of the set and, when the signature gets out of sync due to some problematic update operation, it recomputes the signature from scratch by performing a recovery query. All previous studies \cite{BSS20,MROS,VOS} indicate that, while \vanilla\ exhibits a large update time, it provides good answers to both SE and ACP queries since it can always return the \kminht\ signature of each set.

Our main technical goal is that of preserving the good-quality response for SE and ACP queries as the one guaranteed by the \kminht\ scheme while, at the same time, significantly reducing the update time of the basic \vanilla.
We design a \emph{buffered} version of the \vanilla. Informally, our data structure stores some redundant information useful to reduce the number of necessary recovery queries.

We present a sketch, denoted as $\ell$-\textit{buffered} \kminht, and we rigorously analyze its performances over arbitrary, fully-dynamic streams.  We show that the $\ell$-\textit{buffered} \kminht\  has the following performances (see \Cref{thm:main_result}):
\begin{itemize}
    \item the space used per set is $O(k \log N)$ memory words;
    \item the \textit{amortized} update time\footnote{The \emph{amortized analysis} is a well-known method originally introduced in \cite{Tarjan_amortized} to bound the cost of a sequence of operations, rather than the worst-case cost of an individual operation. This is useful when the worst-case is too pessimistic, as a consequence we average the cost of a \emph{worst case} sequence of operations to obtain a more meaningful cost per operation.} per insertion/deletion is $O(k \log N)$ \Whp
    \item it outputs the \kminht\ signature of any set $A$ in $O(k)$ time. Moreover, such a signature is the same signature that would be computed from scratch from $A$ (and thus the quality of the signature is the one guaranteed by the static \kminht).
\end{itemize}

The above results overall show that it is possible to make the standard static \kminht\ fully dynamic at a reasonable, logarithmic space and amortized update time overheads. This implies that, by paying these small extra costs, the $\ell$-\textit{buffered} \kminht\ allows to get the same accuracy in SE and ACP queries guaranteed by the (static) \kminht. It is worth noticing that our solution works for arbitrary (\ie\ even non-legal) update sequences\footnote{The insertion of an element that already belongs to a set does not change the set itself, as well as the deletion of an element not belonging to the set.}, thus capturing practical scenarios for which the other solutions are not feasible.

We also evaluate our data structure through an intensive set of experiments over both synthetic and real data sets showing that the $\ell$-buffered \kminht\ is a competitive solution in terms of query accuracy and running time performances (see \Cref{sec:experiments}).

\paragraph{Paper's Organization.}
In \Cref{sc:preliminaries}, some instrumental notions and results are given, while in \Cref{sec:comparison} we compare   the $\ell$-\textit{buffered} \kminht\ with the global solution proposed in \cite{BSS20}.
\Cref{sec:ourbufftech} is devoted to the formal definition and the theoretical analysis of the $\ell$-\textit{buffered} \kminht,  while, in \Cref{sec:experiments}, we describe the experimental evaluation we performed on both our solution and that in \cite{BSS20}. In \Cref{sec:conclusion}, we discuss the main open problems. \iffull{Finally, in the Appendix, we provide a description of further previous related work and all the technical details we omitted from \Cref{sec:ourbufftech} and \Cref{sec:experiments}.}{Detailed proofs and supplementary materials are available in the full version of the paper \cite{clementi2024maintainingkminhashsignaturesfullydynamic}.}

\section{Preliminaries}\label{sc:preliminaries}

A popular and effective data-sketch scheme for Jaccard similarity is \textit{MinHashing} \cite{broder1997resemblance,BCFM00}.
Informally, in its original version, a permutation
$h : U \rightarrow U$ is selected uniformly at random, and the sketch assigned to each input set $A \subseteq U$ is  the minimal index among all permuted elements of $A$, \ie\
\begin{equation} \label{eq::minhash}
\minht(A,h) \ = \ \min \{ h(a) \mid a \in A \} \, .
\end{equation}
The key \textit{alignment} property of \minht\ sketches is the fact that they preserve the Jaccard similarity of the original sets in the following rigorous way:
\begin{equation} \label{eq:LSH}
    \Pr[ \minht(A,h) = \, \minht(B,h) ] \ = \  \Jsim(A,B) \, ,
\end{equation}
where the above probability refers to the random choice of function $h$.
Taking advantage of this property, a set $A$ is usually represented as a vector of $k$ independent \minht\ values so that the $i$-th entry of the vector is set to $\minht(A,h_i)$, where $h_i$ is an uniformly and independently chosen hash function. This representation is called \kminht\ \emph{signature}.
Given the signatures of two sets, it is possible to estimate the Jaccard similarity with an additive error of $O(1/\sqrt{k})$ with high probability\footnote{As usual, we say that a sequence of events $\{ \mathcal{E}_k \}_{k \in \mathcal N}$ occurs with high probability if $\Pr[\mathcal{E}_k] \geq 1 - (1/k)^{\Theta(1)}$.} (for short \textit{w.h.p.}) \wrt\ $k$. Moreover, the \kminht\ signatures, combined with the \emph{banding technique} \cite{Leskobook20}, can be used to perform \textit{efficient} ACP, where, from now on, the term \textit{efficient}, refers to the ability to break the $\Omega(m^2)$ \textit{barrier} for the time required to perform ACP tasks for a collection of $m$ sets \cite{Leskobook20}.
A family of functions\footnote{Actually, the solution must also define a probability distribution on $H$: in the sequel, if not specified, the distribution is the uniform one.} $H$ that  satisfies the alignment property (\ie\ \cref{eq:LSH})  is called a \textit{strong Locality Sensitive Hashing} (for short \textit{strong-LSH}) family.
The general notion of LSH families of hash functions can be formalized as follows \cite{indyk1998approximate, HarPeled2012ApproximateNN}.

\begin{definition} \label{def:LSHweak}
    For reals $r_1 > r_2$ and $p_1 > p_2$ in $(0,1)$, a family $\mathcal{H}$ of hash functions is \textit{$(r_1,r_2,p_1,p_2)$-sensitive} if, for any pair of sets $A,B \subseteq U$, the following properties hold: (i) if $\Jsim(A,B) \geq r_1$ then $\Pr[h(A)  =  h(B)] \geq p_1$; (ii) if $\Jsim(A,B) \leq r_2$ then $ \Pr[h(A)  =  h(B)] \leq p_2$.
\end{definition}

After the seminal work in \cite{broder1997resemblance,BCFM00}, MinHashing has been the subject of several studies that have improved the original version along several aspects and parameters (\eg\ \cite{CK07, HHW97, Tho-bottomk-13,LOZ12,dahlgaard2017fast}).
In particular, the expensive use of fully random permutations in MinHashing can be somewhat replaced by weaker notions of hash functions \cite{broder1997resemblance}. In this work, we will make use of the following class of random hash functions introduced in \cite{feigenblat-apx-kminwise-soda11}, for which we restate a result that will be instrumental later.

\begin{definition}[approximate $d$-min-wise independent hash functions] \label{def:apx_min-hash}
Let $0 < \epsilon < 1$ and $2 \leq d \leq N$. A family of hash function $\sH$ on the universe set $U = [N]$ is \textit{$\epsilon$-$d$-min-wise independent} if, for any $A \subseteq U$ and for any $ X \subseteq A $ such that $|X| = d$,
\[  \Pr_{h \in \sH}[\max h(X) \, \leq \, \min h(A \setminus X)] \, = \, (1 \pm  \epsilon) \, \big/ \, \binom{|A|}{|X|} \,  . \]
\end{definition}

\begin{theorem}[\cite{feigenblat-apx-kminwise-soda11}]\label{thm:apx_min-hash_space}
    An $\epsilon$-$d$-min-wise independent hash function $h$ can be stored using $O(d \log \log \frac{1}{\epsilon}+\log \frac{1}{\epsilon})$ space and any computation of $h$ can be performed in time\footnote{Further improvements of such results are shown in \cite{feigenblat2017dk}. However, such new results refer to settings and generalizations that are not useful for our technical purpose.} $O(d \log \frac{1}{\epsilon})$.
\end{theorem}

\section{Comparison with previous global solutions} \label{sec:comparison}

To the best of our knowledge, Bury et al.'s paper \cite{BSS20} is the only analytical work presenting a sketch scheme for fully-dynamic (\ie\ allowing insertions and deletions) streams that allows both SE and ACP queries with provably-good performances.
We will refer to their sketch as BSS.

We remind that the BSS sketch adopts a weaker model than the one assumed in our solution since it does not require recovery queries. On the other hand, BSS can handle legal sequences only. Nevertheless, whenever recovery queries are available \textit{and} the sequences are legal, both solutions are feasible. Hence, in what follows, we provide an analytical comparison under the above assumptions.

Informally, the main idea in \cite{BSS20} is to represent each set $A$ with a sequence of $O(\log N)$ subsets, say $T_i^A$ for $i = 0, \dots, \log{N}$, over a different universe of size $c^2$, where $c$ is a parameter depending on the other parameters as discussed later. We just point out here that the larger is $c$, the better is the quality of the representation, the larger is the used memory and the query time. The main property is that, given two sets $A$ and $B$, there is a suitable index $i$ such that $\Jsim(A,B)$ is roughly equal to $\Jsim(T^A_i,T^B_i)$. The BSS sketch explicitly maintains all such subsets $T^A_i$s by a $O(\log N) \times c^2$ matrix. Any insertion/deletion is implemented by updating a single entry of a suitable row of this matrix (corresponding to a certain $T_i^A$). SE and ACP queries are then answered by first computing the $k$-\minht\ signatures of the $T^A_i$s of the sets involved in the query and then using them in a standard way. This approach allows a fast $O(1)$ update time at the cost of a slower query time of $O(kc^2)$ to compute the signature of a given set $A$. The size of the sketch is $O(c^2 \log N)$ memory words.





As for the approximation quality on similarity estimation, the BSS sketch guarantees the following result.
Given non-negative constants $\delta, r_1, r_2 < 1$ and $\gamma>0$, in \cite{BSS20} it is shown that applying the $1$-\minht\ sketch to their compact representation returns a $(r_1, r_2, (1-\gamma)r_1, 6r_2/(\delta(1-\gamma/5\sqrt{2r_1})))$-sensitive family with probability at least $1-\delta$.
For the sake of comparison, the standard $1$-\minht\ (and hence our solution) guarantees a $(r_1,r_2,r_1,r_2)$-sensitivity family with probability $1$. To answer ACP queries, both schemes need to be twisted by using \kminht\ signatures through the banding technique in order to obtain a better (amplified) sensitivity parameters \cite{Leskobook20}.
However, we point out that the \kminht\ scheme (and hence our solution) has stronger guarantees for two reasons: (i) there exists a sensitive-hashing family for every parameter $r_1 > r_2$ while BSS does not allow every combination of their parameters\footnote{For example, with $r_1 = 0.7$, $r_2 = 0.5$, $\delta = 0.1$ and $\gamma = 0.2$ we have unacceptable parameters $p_1 = 0.56$ and $p_2 = 31.49 > 1 > p_1$. To achieve a legal $p_1 > p_2$ we need $\delta \geq 5.7$ which is still unacceptable.} and, most importantly, (ii) for sufficiently large $k$, it provides a $(r_1,r_2,p_1,p_2)$-sensitive family for every $r_1 > r_2$ and $p_1 > p_2$ with probability $1$, while the BSS scheme always has a failure probability $\delta>0$.

As far as the space is concerned, the BSS sketch uses $O(c^2 \log N)$ memory words per set, where $c^2=O\left(1/(\gamma^4\delta^5 r_1^2)\right)$. We remark that the term $O\left(1/(\gamma^4 \delta^5 r_1^2)\right)$ in the space used by BSS is a constant that gets large values to get reasonable approximation quality\footnote{By selecting $\gamma=0.1$, $\delta=0.1$, $r_1=0.5$ we get a $1/\left(\gamma^4 \delta^5 r_1^2\right) = 4 \times 10^9$.}. \Cref{tab:comparison} summarizes the main features of the known analytical global solutions.

\begin{table*}
\centering
\begin{tabular}{l|cccc}
    \toprule
    Sketch & Space & Update & Query & LSH-sensitive family \\
    \midrule
    \vanilla \cite{BSS20,VOS,MROS} & $O(k)$ & $O(k |A|)$ & $O(k)$ & $(r_1,r_2,r_1,r_2)$\\
    BSS \cite{BSS20} & $O(c^2\log{N})$ & $O(1)$ & $O(c^2k)$ & $(r_1, r_2, (1-\gamma)r_1, 6r_2/(\delta(1-\frac{\gamma\sqrt{2r_1}}{5})))$\\
    $\ell$-buffered \kminht\ & $O(k\log{N})$ & $O^*(k\log{N})$ & $O(k)$ & $(r_1,r_2,r_1,r_2)$\\
    \bottomrule
\end{tabular}
\caption{The table shows the performance bounds of the three known global similarity-sketch solutions for fully-dynamic streaming inputs. Amortized bounds are marked with symbol *.}
\label{tab:comparison}
\end{table*}

Further similarity sketch schemes that support fully-dynamic legal streams but do not provide global solutions \cite{BSS20,VOS,MROS} or adopt different dynamic input models \cite{feigenblat-apx-kminwise-soda11,feigenblat2017dk} are discussed in \iffull{\cref{sec:related}}{the full version \cite{clementi2024maintainingkminhashsignaturesfullydynamic}}.


\section{The \texorpdfstring{$\ell$}{l}-buffered \texorpdfstring{$k$}{k}-\minht } \label{sec:ourbufftech}

In this section, we describe the $\ell$-buffered \kminht\ data structure which is able to maintain a \kminht\ of any subset under an arbitrary sequence of insertions and deletions of elements.

\paragraph{Main Technical Ideas.}
It is well known that the major issue in dynamically maintaining a signature $\kminht(A)$ of a set $A$ is to manage  element deletions.
For example, suppose that you want to delete an element $x \in A$, if $h_i(x)=\minht(A,h_i)$ for some $i\in [k]$, the deletion of $x$ would affect the $i$-th entry of $\kminht(A)$, thus causing a \textit{fault} since the signature gets incomplete.
To restore the signature, the algorithm needs to recover the whole set $A$ via a recovery query, re-hash all the elements and compute the new minimum value resulting in a worst-case $\Theta(|A|)$ additional time.

Our idea is to decrease the rate of such critical  \emph{faults} by making use of a \emph{buffer}: for each hash function $h_i$, instead of keeping track of the minimum hash value only, we store the $\ell$ smallest ones, so that whenever the minimum value is deleted, we can replace it by the second one provided the buffer is not empty, and so on. Intuitively, a large buffer would result in fewer faults but in a space and time overhead.

A simple solution would be to require that the buffer \textit{always} contains the \textit{current} $\ell$ minimum values. This invariant would be too expensive to guarantee. In fact, assume that you want to always keep exactly the $\ell$ minimum values in the buffer. Then any deletion of an element from the buffer would require to compute the minimum among the elements outside the buffer, and thus a recovery query.

Coping with this issue is a bit tricky. As we will see in this section, we need to relax this property of the buffer by maintaining only the hash values smaller than a suitable dynamic threshold $\delta$, thus weakening the invariant property required for the buffer. On one hand, when the buffer is not empty, we show this still guarantees that the elements in the buffer can be used to replace the deletion of the minimum value. On the other hand, the use of the dynamic threshold $\delta \geq 0$ would help us to manage deletions of elements more efficiently by a suitable tuning of the parameter $\delta$. Finally and most importantly, we show our solution will guarantee that the probability that a fault occurs, after the deletions of up to a constant fraction of the elements, is small and it decreases exponentially in $\ell$.

To ensure such a small fault probability, we need to use random-enough hash functions. It turns out that idealized fully-random permutations would work but it is well-known that the use of such functions is too expensive in terms of space usage. As a second technical contribution, we devise a smart way to efficiently apply the property of approximate $d$-min-wise independence \cite{feigenblat-apx-kminwise-soda11} (see \cref{def:apx_min-hash}) in order to get almost the same concentration bound of the one guaranteed by fully-random permutations. Such weaker random hash functions are typically used to amplify the approximation quality of SE responses by using \textit{bottom-$k$ sketches} with only one hash function \cite{feigenblat-apx-kminwise-soda11}, while here we show how to use them to manage fault events due to element deletions in streaming inputs.

\subsection{Definitions and Properties}\label{ssec:datastruct}

We start with some definitions.
Given, a hash function $h$, and a set $A \subseteq U$, we define the following set of pairs: $H(A,h)=\{(h(x),x) \mid x \in A\}$. Moreover, given two elements $x,y \in U$, we say that $(h(x),x ) \le (h(y),y )$ if either $h(x) < h(y)$ or $h(x) = h(y)$ and $x < y$. Finally, we define the following operator.

\begin{definition}[Smallest operator] Given an (ordered) set $X$ and an integer $r>0$, we define $\Smallest(X,r)$ to be the set of the $r$ smallest elements of $X$. If $|X|\le r$, then it returns $X$ itself.
\end{definition}

Our data structure can be formalized as follows.

\begin{definition}[$\ell$-buffered \kminht]
Fix a finite universe $U$ of $N$ elements, two parameters $k, \ell$ in $\mathbb{Z}^+$ and select a vector $\overline{h}=\langle h_1,\dots,h_k \rangle$ where, for each $i=1, \ldots, k$, $h_i: U \to U$ is a hash function. Then, for any subset $A \subseteq U$, an \textit{$\ell$-buffered} \kminht\ of $A$ is a $k$-length vector $\mathcal{S}(A)=\langle S_{1}(A),\dots,S_{k}(A) \rangle$. Each $S_{i}(A)$ is in turn  a tuple $\langle \buff_{A,i}, \delta_{A,i}\rangle$  where $\delta_{A,i} \in (U \cup \{+\infty\}) \times (U \cup \{+\infty\})$ is a \emph{threshold} and $\buff_{A,i} \subseteq H(A,h_i)$ is a set (called $\ell$-buffer) of \emph{at most} $\ell$ pairs of $H(A, h_i)$ having the following \emph{invariants}: for each $i = 1, \dots, k,$
\begin{enumerate}[label=(\roman*)]
    \item \label{list:property1} $\forall x \in A,\; (h_i(x), x) \in \buff_{A,i} \iff (h_i(x),x) \leq \delta_{A,i}$

    \item\label{list:property2} $\vert \buff_{A,i} \vert \leq \ell $

    \item\label{list:property3} $\buff_{A,i} = \emptyset$ if and only if $A = \emptyset$. Moreover, when $A=\emptyset$ we have $\delta_{A,i}=(+\infty,  +\infty)$.
\end{enumerate}
\end{definition}

Notice that \ref{list:property2} allows us to upper bound the size of the data structure, which in turn will affect the performance of our sketch.
Furthermore, observe that, thanks to invariant \ref{list:property1}, for each $i \in \left[ k \right]$, $\buff_{A,i}$ contains the first $|\buff_{A,i}|$ minimum elements of $H(A, h_i)$.
Moreover, invariant \ref{list:property3} ensures that $| \buff_{A,i} |>0$ if $A$ is not empty. As a consequence, $\buff_{A,i}$ always contains a pair $(h_i(x),x)$ such that $h_i(x) = \minh{A,h_i}$ and therefore, given $\struct(A)$, it is always possible to compute the $k$-$\minht$ signature of $A$. \iffull{We provide the detailed pseudo-code of the algorithm that computes the signature in \Cref{alg:getminhash_op}.}{}

\iffull{
\begin{algorithm}[ht]
\caption{\textbello{get\_signature} operation}
\label{alg:getminhash_op}
\begin{algorithmic}[1]

\Procedure{get\_signature}{$\mathcal{S}_A$}
    \For{$i=1, \dots, k$}
        \State $(h_i(x_i), x_i) \leftarrow \min(\buff_{A,i})$
    \EndFor
    \State \Return $(h_1(x_1), \dots, h_k(x_k))$
\EndProcedure

\end{algorithmic}
\end{algorithm}
}{}

Our next goal is to show that the invariants \ref{list:property1}-\ref{list:property3} can be preserved after any update operation on the considered set. In particular, our dynamic data structure $\struct(A)$ supports the following operations.

\begin{enumerate}[label=(\roman*)]
    \item $\textbello{init}(A)$: given a set $A \subseteq U$, it returns $\struct(A)$. Notice that it is not necessary for $A$ to be the empty set.

    \item $\textbello{insert}(\struct(A), x)$: given $\struct(A)$ and an element $x$, it computes the data structure $\struct(A\cup\{x\})$.

    \item $\textbello{delete}(\struct(A), x)$: given $\struct(A)$ and an element $x$, it computes $\struct(A  \setminus \{x\})$.
\end{enumerate}

We now describe how the above operations can be implemented in order to preserve invariants \ref{list:property1}-\ref{list:property3}. \iffull{}{We provide the detailed pseudo-codes of the algorithms in the full version of the paper \cite{clementi2024maintainingkminhashsignaturesfullydynamic}.}
\\

\subparagraph{\textbf{The} \textbello{init} \textbf{operation.}}
We start with the procedure that initialize the data structure starting from a set $A$. \iffull{The pseudo-code of the procedure is given in \Cref{alg:init_op}.}{}
For each hash function $h_i \in \overline{h}$, we simply compute all the hash values of the elements of $A$ and insert into $\buff_{A,i}$ the $\ell$ smallest pairs of $H(A,h_i)$.
Then, set the threshold $\delta_{A,i}$ to $(+\infty, +\infty)$ if $\buff_{A,i}$ is not full (\ie\ $\vert \buff_{A,i} \vert < \ell$), or to the maximum value of $\buff_{A,i}$. Clearly, after the execution of the \textbello{init} procedure, the invariants \ref{list:property1}-\ref{list:property3} hold.\\

\iffull{
\begin{algorithm}[ht]
\caption{\textbello{init} operation}
\label{alg:init_op}
\begin{algorithmic}[1]

\Procedure{init}{$A$}

        \For{$i=1,\dots, k$}

            \State $\buff_{A,i} \leftarrow \Smallest(H(A, h_i), \ell)$
            \If {$\vert \buff_{A,i} \vert = \ell$}
                \State $\delta_{A,i} \leftarrow \max (\buff_{A,i}$)
            \Else
                \State $\delta_{A,i} \leftarrow (+\infty, +\infty) $
            \EndIf
            \State $S_i(A) \leftarrow \langle \buff_{A,i}, \delta_{A,i} \rangle $

        \EndFor
        \State \Return $\mathcal{S}(A)=(S_1(A), \dots, S_k(A))$
\EndProcedure

\end{algorithmic}
\end{algorithm}
}{}

\subparagraph{\textbf{The} \textbello{delete} \textbf{operation.}}
When a deletion of any element $x \in A$ arrives from the input stream, for each $i = 1, \dots, k$, if $(h_i(x),x) \in \buff_{A,i}$ we simply remove $(h_i(x),x)$ from $\buff_{A,i}$.
If after a deletion of an element some $\buff_{A,i}$ becomes empty, we say that a \emph{fault} occurs, and we need to re-initialize through the \textbello{init} operation, the \emph{entire} data structure $\struct(A)$ by accessing to (the current state of) $A$ by a recovery query.
Finally, notice that if the invariants \ref{list:property1}-\ref{list:property3} hold before a delete operation, then they still hold after it. \iffull{The pseudo-code of the procedure is given in \Cref{alg:delete_op}.}{}\\

\iffull{
\begin{algorithm}[ht]
\caption{\textbello{delete} operation}
\label{alg:delete_op}
\begin{algorithmic}[1]

\Procedure{delete}{$\mathcal{S}_A, x$}

    \For{$i = 1, \dots, k$}
        \State $\buff_{A,i} \leftarrow \buff_{A,i} \setminus \{ (h_i(x),x) \}$

        \If{$\buff_{A,i} = \emptyset$}  \Comment{a fault occurs}

            \State $A \leftarrow$ the current version of $A$ by recovery query
            \State $\struct_A \leftarrow \textbello{init}(A)$  \Comment{Initialize $A$'s buffer Algorithm~\ref{alg:init_op}}

            \State \Return

        \EndIf
    \EndFor

\EndProcedure

\end{algorithmic}
\end{algorithm}
}{}

\subparagraph{\textbf{The} \textbello{insert} \textbf{operation.}}
Let $x \in U$ be the new element we add to $A$.
Fix an index $i \in \left[k \right]$.
When $(h_i(x),x) > \delta_{A,i}$ we do nothing, and the invariants are clearly preserved.
Otherwise, $(h_i(x), x) \leq \delta_{A, i}$, and we add the pair $(h_i(x), x)$ to $\buff_{A,i}$. There are two cases. If $\vert \buff_{A,i} \vert < \ell$, then we keep the threshold $\delta_{A,i}$ unchanged. Notice that the invariants are preserved.
In the second case $\buff_{A,i}$ has either $\ell$ or $\ell+1$ elements. We restore the invariant~\ref{list:property2} by removing the maximum element from the buffer if $\vert \buff_{A,i} \vert = \ell+1$. Finally, we restore invariant~\ref{list:property1} by setting the threshold $\delta_{A, i}$ to the maximum element currently in the buffer $\buff_{A,i}$.
\iffull{The pseudo-code of the procedure is given in \Cref{alg:insert_op}.}{}\\

\iffull{
\begin{algorithm}[ht]
\caption{\textbello{insert} operation}
\label{alg:insert_op}
\begin{algorithmic}[1]

\Procedure{insert}{$\mathcal{S}_A, x$}
    \For{$i = 1, \dots, k$}
        \If{$(h_i(x), x) \leq \delta_{A,i}$}
            \State $\buff_{A,i} \leftarrow \Smallest(\buff_{A,i} \cup \{ (h_i(x), x) \},\; \ell)$
            \If{$\vert \buff_{A,i} \vert = \ell$}
                \State $\delta_{A,i} \leftarrow \max (\buff_{A,i})$  \label{alg:line:update_delta}
            \EndIf
        \EndIf
    \EndFor
\EndProcedure

\end{algorithmic}
\end{algorithm}
}{}

\noindent A reader can now understand the role of the thresholds $\delta_{A,i}$. On the one hand, it allows us to weaken the requirement of keeping always the $\ell$ minimum hash values of $h_i(A)$ in $\buff_{A,i}$ (that cannot be guaranteed efficiently). On the other hand, by keeping in $\buff_{A,i}$ exactly the elements below the threshold $\delta_{A,i}$, we guarantee that any discarded element cannot ever become the minimum hash value $\minh{A,h_i}$ until a fault occurs.
Observe that any naive strategy, in which a delete operation is managed by removing $(h_i(x),x)$ from $\buff_{A,i}$, while inserting an element $x$ is implemented by simply inserting $(h_i(x),x)$ into $\buff_{A,i}$ (if there is enough space), might produce a configuration of the data structure from which it is not possible to recover the min-hash signature of the set.
\iffull{
An instance of this scenario is provided in \Cref{fig:counter_example}.

\begin{figure}[ht]
    \centering
    \includegraphics[width=\linewidth]{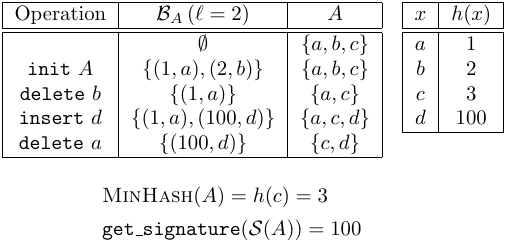}
    \caption{A counterexample showing that the naive strategy might produce a configuration of the data structure from which it is not possible to recover the min-hash signature of the set. The naive strategy: a delete operation is managed by removing the element from the buffer, while an insertion is implemented by simply inserting the element in the buffer (if there is enough space). In the example, a sequence of update operations is shown for $k=1$ and $\ell=2$. After such a sequence, a query is performed in order to obtain the \minht\ signature of the set but the value returned by the data structure is wrong since it does not coincide the the \minht\ of the set.}
    \Description{...}
    \label{fig:counter_example}
\end{figure}

}{
See the full version \cite{clementi2024maintainingkminhashsignaturesfullydynamic} for an instance of this scenario.
}

Finally, notice that our data structure works also with non-legal streams.
Indeed, we store in the buffers the hash values with the corresponding elements, and thus any insertion of an element which is already in the set, or a deletion of an element which does not belong to the set, does not change at all the buffers.

\subsection{Performance Analysis over Fully-Dynamic Data Streams} \label{ssec:analysis}

We are now ready to provide the main theoretical result of this paper. We essentially show that setting $\ell = \Theta(\log{N})$ and sampling the hash functions $\overline{h}$ from a family of $(\varepsilon, \ell)$-min-wise independent hash functions with $\varepsilon = O(1)$ is sufficient to ensure that our $\ell$-buffered $k$-\minht\ guarantees an amortized cost of $O(k\log{N})$ per operation, \Whp


We fix $k$, $\ell$ and sample u.a.r. $k$ hash functions $h_1,\dots, h_k$ from a family $\mathcal{H}$ of $(O(1), \ell)$-min-wise independent hash functions (see \cref{def:apx_min-hash}).

To analyze the performance of our data structure we have to provide some implementation details. We maintain each $\buff_{A,i}$ through a balanced binary search tree (\eg\ Red-Black tree \cite{cormen01introduction}) of at most $\ell$ elements, which allows us to insert, delete and search for an element in $O(\log \ell)$ time. This data structure is then able to return the minimum (maximum) stored element in constant time.

We first study the space usage of our sketch.

\begin{lemma}\label{lm:space_usage}
    The overall memory space required by the $\ell$-buffered \kminht\ of  all $m$ sets and the hash-function vector $\overline{h}$ is $O(m k \ell)$.
\end{lemma}

\begin{proof}
    Each set $A_j$ (for $j=1,\dots,m$) is maintained by using the $\ell$-buffered \kminht: this requires $O(k \ell)$ memory words since each $\buff_{A_j,i}$ needs $O(\ell)$ space. Moreover, thanks to \Cref{thm:apx_min-hash_space} and since $\varepsilon = O(1)$, we can store all the hash functions in $O(k\ell\log ( \frac{1}{\varepsilon}))=O(k \ell)$.
\end{proof}

We now bound the running time of each operation. Let $T_h$ be the time needed to compute the hash value of an element $x$ for any hash function in $\mathcal{H}$. The following lemmas bound the time complexity of the \textbello{insert} and the \textbello{init} operation, respectively. Notice that to perform the insertion of an element $x$ into a set $A$, we have to compute $k$ hash values and update the corresponding $\buff_{A,i}$s. Such an update requires a constant number of operations on the associate Red-Black tree, each of them taking $O(\log \ell)$ time. Moreover, we observe that the time required by the \textbello{init} operation is upper bounded by $|A|$ times the time required by an insertion. Therefore, we have the following.

\begin{lemma}
    \label{lm:time_init}
    The running time of any \textbello{init} for a set $A$ is $O(|A| \cdot ( k\log \ell + kT_h))$.
\end{lemma}

\begin{lemma}
    \label{lm:time_insert}
    The running time of any \textbello{insert} operation on any set $A$  is $O(k\log \ell + kT_h)$.
\end{lemma}

As for the \textbello{delete} operations, we first observe that, in the worst-case, we need to re-execute an \textbello{init} operation: this implies a running time $O(|A| \cdot (k\log{\ell} + k T_h))$. However, as we will show via a refined analysis, the amortized bound is much better.

We start the amortized analysis by first showing that the probability of having a fault, when a set $C \subseteq A$ of elements is deleted, decreases exponentially in $\ell$.

\begin{lemma}\label{thm:whp}
Consider a set $A \subseteq U$ of size at least $\ell$, and any subset $C \subset A$.
Let $\,\Smallest(H(A,h),\ell)\,$ be the set of the $\ell$ smallest pairs $(h(x),x)$, where $x \in A$ and $h$ is an hash function chosen uniformly at random from a family $\sH$ of $(\varepsilon,\ell)$-minwise independent hash functions.
Then, the probability of the event $\mathcal{F} = \{ \Smallest(H(A,h),\ell) \subseteq H(C,h)\}$ is at most $(1+\varepsilon)\left( \frac{|C|}{|A|} \right)^\ell$.
\end{lemma}

\iffull{
\begin{proof}
If $|C| < \ell$ then $P(\mathcal{F}) = 0$.
Now consider the case $|C| \geq \ell$.
Let $C' \subseteq C \subset A$ be any subset of size $\ell$.
From \cref{def:apx_min-hash}, we get
\begin{align*}
&\Pr_{h \in \sH}[ H(C',h) = \Smallest(H(A,h),\ell) ] \, =\\ \,
&\Pr_{h \in \sH}[ \max h(C') \leq \min h(A \setminus C')]
\, \leq
\, (1+\varepsilon)\frac{1}{\binom{|A|}{\ell}} \, .
\end{align*}
By applying the union bound over all $C' \subseteq C$ of size $\ell$, the probability that \textit{any} event $\mathcal{F}$ does happen can be bounded as follows:

\begin{align*}
\Pr_{h \in \sH}[ \mathcal{F}]
&= \Pr_{h \in \sH}[\exists C' \subseteq C : |C'| = \ell \land H(C',h) = \Smallest(H(A,h),\ell)]\\
&\leq (1+\varepsilon)\frac{\binom{|C|}{\ell}}{\binom{|A|}{\ell}}
= (1+\varepsilon)\frac{|C|!}{\ell! \cdot (|C| - \ell)!} \frac{\ell! \cdot (|A| - \ell)!}{|A|!}\\
&= (1+\varepsilon)\frac{|C| (|C| - 1) \dots (|C| - \ell + 1)}{|A| (|A| - 1) \dots (|A| - \ell + 1)}
\leq (1+\varepsilon)\left( \frac{|C|}{|A|} \right)^{\ell}
\end{align*}
\end{proof}
}{}

\noindent Now, we first observe that the probability that a fault occurs within a sequence of element insertions/deletions is maximized when \textit{all} the operations are deletions. Therefore, we have the following.

\begin{corollary}
    \label{cor:prob_fault}
    Let $A \subseteq U$ be a set of size $n \ge \ell$, and let $\struct(A)$ be the $\ell$-buffered \kminht\ sketch of $A$ after a \textbello{init} operation.
    Consider any sequence of insert and delete operations.
    There exists a constant $\gamma > 1$ such that if $\ell \geq \gamma \log n$ then the probability that a fault occurs before $\frac{n}{4}$ operations\footnote{We fixed the  fraction $n/4$ just to avoid the use of other parameters: any value $n/b$ with a constant $b > 1$ would work.} is at most $\frac{1}{n^5}$.
\end{corollary}

The previous corollary states that, starting with a full buffer, with high probability, $\Omega(|A|)$ deletions must be executed before a fault occurs. This can be used to amortize the cost of a fault.

\begin{lemma}\label{thm:ammortized}
Let $\ell = \Theta(\log{N})$, and $A \subseteq U$ be any set of elements. Moreover, let $\struct(A)$ be the $\ell$-buffered $k$-\minht\ sketch computed by the \textbello{init} operation.
In any arbitrary sequence of update operations on $\struct(A)$,  each update operation takes $O(k\log\log{N} + k T_h)$ amortized time \Whp (\wrt\ $|A|$).
\end{lemma}

\iffull{
\begin{proof}
Since each fault triggers a new \textbello{init} operation that restores the assumption of the claim, we focus only on a sequence of insert/delete operations between the first \textbello{init} to the first fault.

We use the accounting method \cite{Tarjan_amortized}. The idea is to pay the cost of the fault by charging it to the previous insert and delete operations. More precisely, we assign \emph{credits} to each insert and delete operation that we will use to pay the cost of the fault. Formally, the \emph{amortized} cost of an insertion or of a fault-free delete operation is defined as the \emph{actual} cost of the operation, plus the credits we assign to it. While the amortized cost of the fault is defined as its actual cost minus the credits (accumulated from the previous operations) we spend to pay for it. We need to carefully define such credits in order to guarantee that the sum of the amortized costs is an upper bound to sum of the actual costs.

Let $\Delta = \alpha \cdot (k\log{\ell} + kT_h)$, for some large enough constant $\alpha$.
By construction and for \Cref{lm:time_insert}, the \textbello{insert} operation always has an \emph{actual} cost of at most $\Delta$. 
When a fault does not occur, the actual cost of \textbello{delete} is at most $\Delta$ as well.
We now assign a credit of $3\Delta$ to each insert/delete operation in the sequence before the fault, thus obtaining an amortized cost of $4\Delta = O(k\log \log N + kT_h)$.

Now, let $n$ be the size of $A$ at the beginning of the sequence of the operations, $n'$ be the size of $A$ when the fault occurs, $n_D$ and $n_I$ be the number of \textbello{delete} and \textbello{insert} operations executed in this sequence, respectively.
Notice that $n' = n - n_D + n_I$.
The actual cost of the last operation (the fault) is at most $n' \Delta$ due to the \textbello{init} operation (\Cref{lm:time_init}), while the amount of accumulated credits is $(n_D + n_I) 3\Delta$.
Thus the amortized cost $\hat{c}(\text{fault})$ of the fault is
\begin{align*}
    \hat{c}(\text{fault})
     & \leq n'\Delta - (n_D+n_I)3\Delta
    = (n + n_I - n_D)\Delta - (n_D+n_I)3 \Delta\\
    & = n\Delta -4n_D\Delta -2n_I\Delta \leq n \Delta - 4 n_D \Delta \, .
\end{align*}
Since a fault occurs, then $n_D \geq \ell$.
Now, if $n < 4 \ell$ clearly $\Hat{c}(\text{fault}) < 0$ with probability $1$.
Otherwise, by \cref{cor:prob_fault}, since $\ell = \Theta(\log{N})$, we have $n_D \geq n/4$ with probability at least $1-n^{-5}$. Hence, $\Hat{c}(\text{fault}) \leq 0$, \Whp
Finally, observe that the $3(n_I+n_D)\Delta$ credits we are using to pay the actual cost of the fault are from \emph{new} insert/delete operations, which implies that such credits are spent at most once. Therefore,
summing up all amortized costs, we can upper bound the total cost of the entire sequence.
\end{proof}
}{}

Finally, combining \Cref{thm:ammortized}, \Cref{lm:space_usage}, and \Cref{thm:apx_min-hash_space}, we obtain our main result.

\begin{theorem}\label{thm:main_result}
    The $\ell$-buffered \kminht\ data structure
    is able to maintain the \kminht\ signatures of any fully-dynamic set collection of $m$ sets over a universe $U$ of size $N$ with the following performance guarantees:
    \begin{itemize}
        \item Memory usage: $O(k \log N)$ memory words per set;

        \item Time: any sequence of update operations requires $O(k \log{N})$ amortized time per operation with high probability \wrt\ the maximum size of the sets in the collection;

        \item Response Quality: it can return the $k$-\minht\ signature of each set of the collection in time $O(k)$.
    \end{itemize}

\end{theorem}




\section{Experimental Evaluation} \label{sec:experiments}

We complement our theoretical analysis with an extensive set of experiments performed on both synthetic and real data sets. We remark that the former are suitably chosen in order to stress our solution.

In \Cref{subsection:params}, we first estimate the running time of our sketch, achieving a speedup up to $745$x \wrt\ \vanilla\ on fully-dynamic data stream (see \Cref{tab:speed_up}), by paying some slightly-larger memory footprint (consistently with our theoretical analysis).\footnote{If instead one insists on having the exact same memory, the quality of the Jaccard similarity estimation of our solution is just a bit worse than the one obtained by \vanilla\ (see \Cref{fig:SE_test}).}

We then compare our solution with the BSS sketch in \Cref{sec:comparison_BSS_exp}: the latter has a really fast update time, while a quite large query time. We show that in a sequence of interleaved update and query operations, our solution significantly outperforms the BSS sketch as soon as the rate of the query operations is above the threshold of 5-10\% of the overall operations (see \Cref{fig:time_update_query}). To cope with the slow query time, the authors of \cite{BSS20} propose in the same paper a variant of their sketch called \emph{BSS-proactive} that has a slower update time and a faster query time. We also implemented this variant which results to exhibit a similar time performance as our solution (see \Cref{fig:time_comparison}). On the other hand, the major advantage of our sketch relies on the quality of the Jaccard similarity evaluation: we show that under the RMSE, we are a order of magnitude more precise than both BSS sketches (see \Cref{fig:SE_test}). Furthermore, our solution exhibits a much smaller variance.
Such a quality improvement on Jaccard similarity evaluation is also reflected on ACP queries on real datasets.
\iffull{The discussion of this experiment can be found in \Cref{apx:quality_ACP}.}{More details and further experiments are provided in the full version of the paper \cite{clementi2024maintainingkminhashsignaturesfullydynamic}.}

In this section, we will refer to our $\ell$-buffered \kminht\ as BMH.



\paragraph{Implementation Details.} We implemented BMH and BSS in \textbello{C++} and compiled with GCC version 10.2.1 using flags -O3, -fopenmp and -mavx
\footnote{We actually implemented a serial code. In the experiments aiming at measuring the execution time, no parallelism is employed, while we run independent experiments in parallel in all the other cases to speed up the execution.}.
All the experiments were run on a machine with 2.3 GHz Intel Xeon Gold 5118 CPU with 24 cores, 192 GB of RAM, cache L1 32KB, shared L3 of 16MB and UMA architecture.

Since the state-of-the-art solutions only work on \emph{legal} streams, we implemented a simplified version of our BMH sketch that works for \emph{legal} streams only.
Hence, each $\ell$-buffer stores the hash values only.
Moreover, since in our experiments the value of $\ell$ is reasonably small, instead of using a balanced binary search tree, we implement the $\ell$-buffers as unordered arrays and explicitly maintain the minimum value for each buffer. Notice that this would not affect the theoretical asymptotic bounds stated in \Cref{thm:main_result}.

We implemented both BSS and BSS-proactive sketches\footnote{No actual implementation is available.}, where we fix the parameter $\alpha = 0.1$, according to the setting in \cite{BSS20, MROS}.
The BSS-proactive differs from the BSS sketch by dynamically maintaining the \kminht\ signatures as in our BMH sketch, instead of computing it from scratch at query time.
We notice that the BSS-proactive suffers of the same problem of \minht, \ie\ a deletion may cause a fault.
Observe that in the BSS-proactive a deletion may cause a fault.

For the sake of a fair comparison, we use the same hash functions to compute the signatures. In particular, we use \textit{Tabulation Hashing} \cite{tabulation,Zobrist1990ANH} with $8$ tables each of $16$ entries to compute signatures.

\iffull{A more detailed description of the implementations is provided in \Cref{apx:implementation}.}{}
Our implementation is publicly available at \url{https://github.com/Alessandrostr95/FullyDynamicKMinHash}.

\paragraph{Datasets.}
We use two types of synthetic datasets. In the first one, that we use to measure the execution time of the considered solutions, we simply sample uniformly at random subsets of different sizes from the universe $U = \left[ 2^{32} \right]$, and store them with a sparse representation.

To compare the precision of the Jaccard similarity estimation obtained by the different sketches, we generate a synthetic dataset following the benchmark of \cite{CDJ+00}.
In more detail, we generate a set $A$ by sampling elements from the universe $U$ with probability $q = 0.05$.
For each set $A$ we further generate a second set $A'$ as follow: each element $x \in A$ is inserted into $A'$ with probability $p_1$, while each element $x \not\in A$ is inserted with probability $p_2$.
We choose $p_1,p_2$ such that i) the size of $A'$ is almost the same size of $A$ and ii) the Jaccard similarity $\Jsim(A,A')$ is, in expectation, a given parameter $J$.

As in \cite{VOS,pang2024similarity, Plos18}, we evaluate ACP queries among neighborhoods of nodes in real networks \cite{LiveJournal1,LiveJournal2,YouTube-Orkut} (more details are provided in \iffull{\Cref{apx:datasets_description}}{the full version}).

\subsection{Running Time Evaluation} \label{subsection:params}

The first set of experiments is devoted to empirically evaluate the time bounds proved in \Cref{sec:ourbufftech} of our data structure over streams of operations. To do so, we measure the execution time needed to process sequences of operations with special attention to the impact of the buffer size $\ell$ on the performances.
Since the query time to obtain a signature is independent of $\ell$ and the size $n$ of the set, we consider only sequences of update operations.
\Cref{fig:time_ell} shows the average execution time and the average number of faults following a series of updates, for different values of the parameters $\ell$ and $k$.

\Cref{fig:test_ell_medium_faults} shows that the average number of faults decreases exponentially in $\ell$ as predicted by our theoretical analysis.
Even though our theoretical analysis requires $O(\log{n})$-min-wise independent hash functions to guarantee a low fault rate, these experiments empirically suggest that Tabulation Hashing can be enough for practical purposes. 

The observed execution times are reported in \Cref{fig:test_ell_medium}.
We can see that the best running time is obtained when $\ell$ is set to a value close to $\log_2{n}$ which is in line with our analysis, while there is no advantage to use larger values of $\ell$.
As a result, since in real applications the size of $A$ is unknown a priori, or simply changes over time, from now on we set $\ell = 32$ (\ie\ $\ell = \log_2{|U|}$).

Finally, in \Cref{tab:speed_up} we report the comparison with \vanilla.
It is worth noticing that our solution strongly outperforms \vanilla, gaining a speed-up up to $745$x. This is a clear evidence that the buffering technique results in a huge running time improvement at a reasonable cost of extra space.
\iffull{A more detailed description of the implementations is provided in \Cref{apx:running_time_evaulation}.}{}


\begin{figure}
\begin{subfigure}{.23\textwidth}
  \centering
  \includegraphics[width=\linewidth]{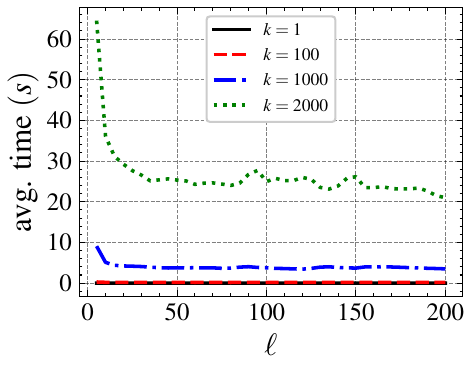}
  \caption{}
  \label{fig:test_ell_medium}
\end{subfigure}
\begin{subfigure}{.23\textwidth}
  \centering
  \includegraphics[width=\linewidth]{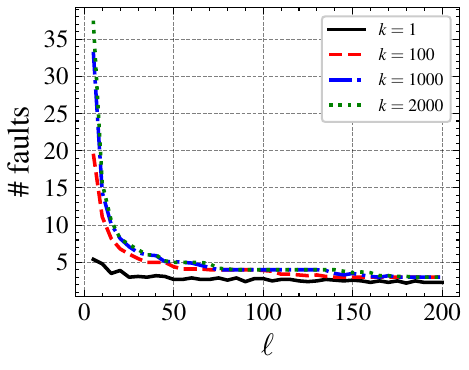}
  \caption{}
  \label{fig:test_ell_medium_faults}
\end{subfigure}
\caption{(a) shows the average time of computing $n$ inserts followed by $n$ deletions (with $n = 2^{16}$) for different values of $k$, varying the size of the buffer $\ell$ from $5$ to $200$.
(b) shows the average number of faults.
For readability reason the case $\ell = 1$ is omitted.
}
\Description{...}
\label{fig:time_ell}
\end{figure}

\begin{table}
    \centering
    \begin{tabular}{c|ccc}
    \toprule
    $n$ & \vanilla & BMH & speedup \\
    \midrule
    $2^{12}$ & $96.376s$ & $0.404s$ & $238$x \\
    $2^{16}$ & $2049.953s$ & $2.966s$ & $691$x \\
    $2^{19}$ & $15279.49s$ & $20.484s$ & $745$x \\
    \bottomrule
    \end{tabular}
    \caption{Average running time comparison in executing $2n$ operations between \vanilla\ and BMH with $\ell = 32$, for $k=2000$ hash functions.}
    \Description{...}
    \label{tab:speed_up}
\end{table}

\subsection{Comparison with BSS Sketches}\label{sec:comparison_BSS_exp}
We now compare the BMH sketch to the BSS one (and its proactive version).
We first compare the execution time of the sketches over sequences of operations, then we turn our attention to the approximation quality achieved on the Jaccard similarity estimation.
For the sake of fairness, in all experiments described in this section, we ensure that all sketches always use the same memory size.
Since the memory words required for our sketch is $O(k\log{|U|})$ and for the BSS sketches is $O(c^2\log{|U|})$, we set $c^2 = k$.



\paragraph{Execution Time.}
We perform three types of experiments.
The first type consists in measuring the execution time obtained by BSS, BSS-proactive, and our BMH sketch to perform a sequence of $n$ insertions and $n$ deletions, varying the memory size.
In \Cref{fig:time_update}, for each value of the memory size, we report the average execution time of $100$ independent experiments.
We can see that the theoretical behaviour of BSS fits the experiment results, since its update time is constant and independent from the memory size, while the running time of our BMH and BSS-proactive sketches grows up as the memory increases due to the overhead to compute hash values and to deal with the faults.
The reason why BSS-proactive is slower than our BMH sketch is the higher number of faults that occurs in the former.
Although our faults are more expensive we guarantee that the number of faults is nearly constant (see \Cref{fig:test_ell_medium_faults}), while they have no guarantees on the number of faults.


In the second type of experiments we focus on the evaluation of the query time to obtain the signature of a set. We compare the execution times of the three sketches over a sequences of $n$ queries on sets of $n$ elements\footnote{Since all the sketches returns a \kminht\, the running time for computing the Jaccard similarity, given the signatures, is the same. Hence, what really differentiates the sketches is the runtime for computing the \kminht\ signature.}. The average execution time over $100$ experiments is reported in \Cref{fig:time_query}.
Here the results present an opposite behavior compared to the previous ones.
The BSS-proactive and our BMH sketches have a nearly constant time, since they precompute the \kminht\ signature and just return a pointer to the signature.
Meanwhile BSS has to compute it from scratch incurring in a $O(kc^2)$ complexity.

\begin{figure}
\begin{subfigure}{.15\textwidth}
  \centering
  \includegraphics[width=\linewidth]{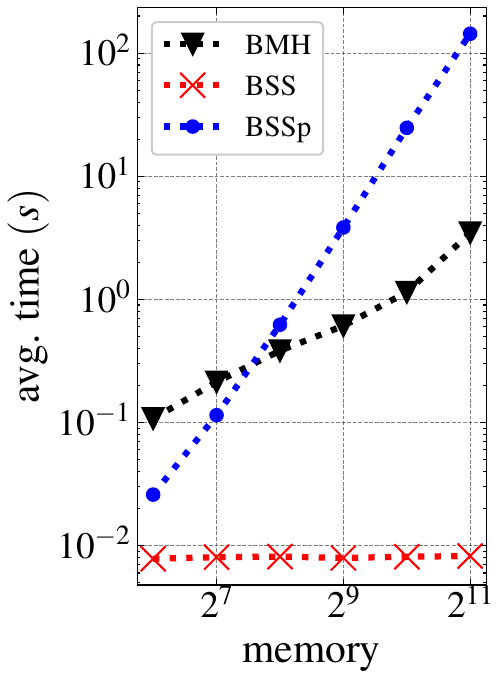}
  \caption{$n = 2^{16}$ updates.}
  \label{fig:time_update}
\end{subfigure}%
\begin{subfigure}{.15\textwidth}
  \centering
  \includegraphics[width=\linewidth]{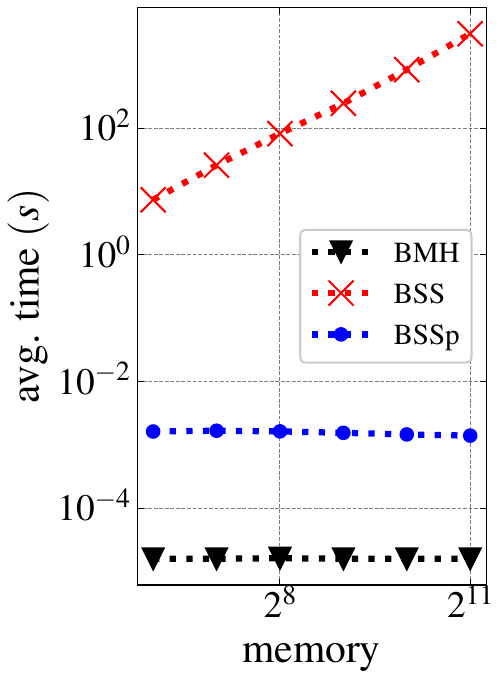}
  \caption{$n = 2^{16}$ queries.}
  \label{fig:time_query}
\end{subfigure}
\begin{subfigure}{.15\textwidth}
  \centering
  \includegraphics[width=.95\linewidth]{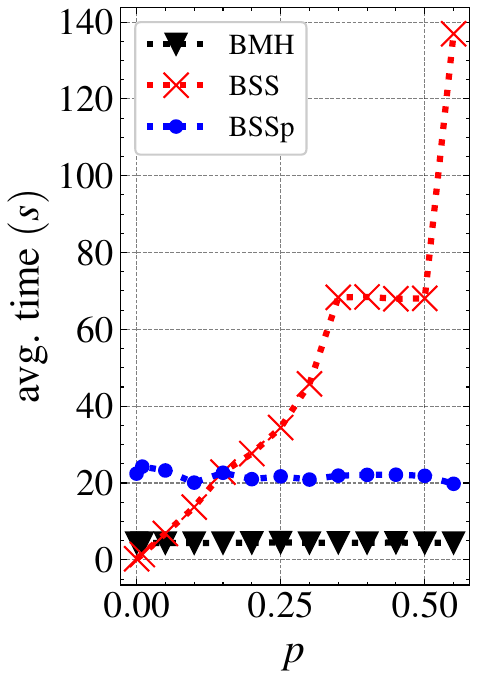}
  \caption{$n = 2^{17}$ operations.}
  \label{fig:time_update_query}
\end{subfigure}

\caption{(a) comparison as memory changes of $n = 2^{16}$ insertions followed by $n$ deletions.
(b) comparison as memory changes of $n=2^{16}$ queries.
The $x$-axes indicates the value $k$ and $c^2$. Notice the log-log scale.
(c) time comparison for a sequence of $n$ operations varying the fraction $p$ of query, for fixed memory $k = c^2 = 1024$ and $\ell = \log |U| = 17$.
}
\Description{...}
\label{fig:time_comparison}
\end{figure}

The previous two types of experiments show that the our BMH is slower than BSS for updates while it is faster for queries. Moreover, our BMH always favorably compares with BSS-proactive for both queries and updates.
In the third type of experiments, both queries and updates are taken into account.
We fix $k = c^2 = 1024$ and execute $n$ operations, where a fraction $p$ are queries and the remaining $1-p$ fraction are updates\footnote{We set $c^2=1024$ as discussed in \cite{BSS20}.}.
\Cref{fig:time_update_query} shows that even when $p$ is small (about $5\%$ of total operations), our BMH is significantly faster than BSS.
The theoretical analysis also states that the running time of BSS deteriorates as the number of queries growths which is confirmed in practice. Meanwhile, the BMH and BSS-proactive's running time is almost unaffected by the type of operations performed (unless they are almost only queries).


Finally, we  point out that in \cite{BSS20}, the authors claim that \vanilla\ and BSS-proactive are comparable as far as the update time is concerned. This is in contrast with the outcomes of our experiments where \vanilla\ is much slower than any other solution. This is due to the restricted type of sequences of updates employed in \cite{BSS20}\footnote{
After inserting an element, they \emph{immediately} delete it with probability $\frac{1}{10}$.
Such a regular deletion procedure benefits both BSS-proactive and \vanilla\ due to the low probability of a fault.
We want to underline that if we had used the setting in \cite{BSS20}, even with $\ell = 2$, no fault would have occurred for our BMH sketch.}.

\paragraph{Quality Comparison.}\label{paragraph:quality_comparison}
We now turn our attention to the approximation quality achieved on the similarity estimation by measuring the Root Mean Square Error (RMSE) of the computed Jaccard similarities. We compare the BMH, BSS and \vanilla\ sketches using the same fixed memory size.

We use the dataset of \cite{CDJ+00} as explained before.
For different values of $J$ we
generate $1000$ pairs of sets such that their expected Jaccard similarity is $J$. For each pair of sets, we estimate the Jaccard similarity by using each sketch and compute the exact Jaccard similarity.


In \Cref{fig:SE_test} we report the RMSE for all the three sketches as well as the standard deviation. It can be clearly seen that our BMH sketch is much more accurate and stable than the BSS sketch for every value of $J$. Indeed, our BMH obtains an error $10$ times smaller than the error achieved by the BSS. We think that the main reason for that relies on the choice of $c^2$. As we already discussed, the actual value chosen in the experiments is far from the one the theory would require to provided a good error bound.
Furthermore, the large standard deviation resulting from the BSS sketch is due to the probability $\delta$ of failure in reporting a bounded approximation of the Jaccard similarity. \iffull{More experiments in \Cref{apx:quality_SE}.}{}

Finally, we point out that the error obtained by our BMH is comparable with the one achieved by \vanilla. This suggests that the memory overhead we pay for having fast update time does not significantly affect the quality of the estimation.

\begin{figure}[ht]
    \centering
    \includegraphics[width=0.5\linewidth]{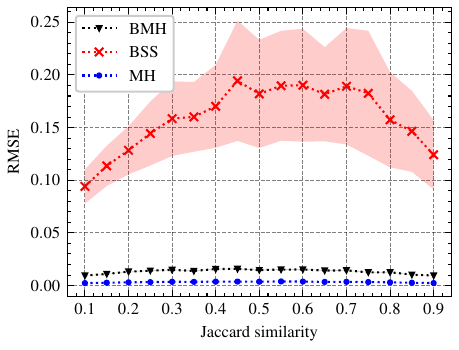}
    \caption{RMSE for different values of Jaccard similarity, from $0.1$ to $0.9$. The size of the sketches is $1024 \times \log{\vert U \vert}$ memory words, where $|U| = 2^{17}$. Thus \vanilla\ (MH in the figure) uses  $k \times \log{\vert U \vert} = 1024 \times 17$ hash functions.
    The size of the sets are $\approx 6500$.
    Each point is the RMSE for $1000$ experiments.
    The shaded areas represent the standard deviation of the experiments.}
    \Description{...}
    \label{fig:SE_test}
\end{figure}

\iffull{
\subsection{Quality Comparison on ACP Queries}\label{apx:quality_ACP}
To complement our experiments on synthetic datasets, we evaluate ACP queries on the real datasets described in \Cref{apx:datasets_description}.
For each dataset, we selected a similarity threshold $J$ to determine the pairs to report. Pairs with a similarity above $J$ are considered \emph{True Positives} (TP), while non-reported pairs with a similarity below $J$ are considered \emph{True Negatives} (TN). Conversely, reported pairs with a similarity below $J$ are \emph{False Positives} (FP), and non-reported pairs with a similarity above $J$ are \emph{False Negatives} (FN).

We compare our BMH sketch with the BSS sketch according to three different metrics:

\begin{align*}
    &\mbox{Precision} \, = \,  \frac{TP}{TP+FP}, \\ &\mbox{Recall}  \, = \,  \frac{TP}{TP+FN}, \\
   & F_1 \, = \, 2 \cdot \frac{\mbox{Precision} \, \times \, \mbox{Recall}}{\mbox{Precision} \, + \, \mbox{Recall}} \, .
\end{align*}

Consistently with our previous experiments, we maintain the same memory footprint for both sketches and the same number of hash functions for LSH.
In particular, we use no more than $2100$ hash functions, and we set the the number of bands $b$ and the size $r$ of each band in such a way to obtain a $(r_1, r_2, p_1, p_2)$-sensitive family with $p_1 = 0.8$.
For the BSS sketch, we selected $b$ and $r$ to achieve the best performance (\ie\ higher $precision$, $recall$ and $F_1$) while maintaining the same number of hash functions.
This empirical process for selecting the BSS best parameters is necessary since no helpful closed-form formula  is available for this task.

Concerning the threshold $J$ used for ACP queries, we select the one such that the number of effective pairs is at most $4\%$ of the total possible pairs.
Moreover, the different datasets allow us to select different values of $J$ stressing the LSH algorithm.

\Cref{tab:APC_results} presents all the parameters selected and the results of ACP queries for each sketch, across the different datasets.

The results of the experiments highlight that both sketches return a number of pairs comparable with the number of effective pairs, although our BMH sketch obtains higher Recall values.

We point out that when the number of returned pairs is in the same order of magnitude as the number of effective pairs, having a high Recall is more convenient than having a high Precision since a fast post-processing can be performed to filter the returned pairs and thus enhancing the Precision.
Conversely, improving Recall solely from the returned pairs is not feasible, as it would require to reconsider all potential pairs.

We also observe that for the Orkut dataset, the BSS sketch achieves a significantly better Precision than BMH. This is essentially due to the fact that it returns much fewer pairs than the effective ones.

Finally, we emphasize that while for \minht\ there is an analytical way to set $b$ and $r$, it is unclear how these parameters need to be set in the BSS in order to obtain good Precision/Recall scores.

\begin{table*}[ht!]
    \centering
    \begin{tabular}{llc|ccc|ccc|cc}
    \toprule
         && Sketch & $J$ & $b$ & $r$ & Precision & Recall & $F_1$ & \makecell{Effective \\ Pairs} & \makecell{Returned \\ Pairs} \\
    \midrule
        \parbox[t]{2mm}{\multirow{6}{*}{\rotatebox[origin=c]{90}{$d=1$}}}

        &\multirow{2}{*}{\textbello{LiveJournal}}
        & BMH & \multirow{2}{*}{$0.1$} & $700$ & $3$ & \textbf{0.808} & \textbf{0.955} & \textbf{0.875} & \multirow{2}{*}{$113,117$} & $133,671$  \\
        && BSS & & $420$ & $5$ & $0.439$ & $0.753$ & $0.555$ & & $193,873$ \\

    \cmidrule{2-11}
        &\multirow{2}{*}{\textbello{Orkut}}
        & BMH & \multirow{2}{*}{$0.1$} & $700$ & $3$ & $0.231$ & \textbf{0.752} & $0.354$ & \multirow{2}{*}{$12,604$} & $40,963$ \\
        && BSS & & $350$ & $6$ & \textbf{0.74} & $0.4$ & \textbf{0.52} & & $6,814$\\

    \cmidrule{2-11}
        &\multirow{2}{*}{\textbello{YouTube}}
        & BMH & \multirow{2}{*}{$0.05$} & $400$ & $2$ & \textbf{0.247} &\textbf{0.832} & \textbf{0.382} & \multirow{2}{*}{$23,763$} & $79,852$\\
        && BSS & & $160$ & $5$ & $0.13$ & $0.30$ & $0.18$& & $53,439$\\

    \midrule[.66pt]

        \parbox[t]{2mm}{\multirow{6}{*}{\rotatebox[origin=c]{90}{$d=2$}}}

        &\multirow{2}{*}{\textbello{LiveJournal}}
        & BMH & \multirow{2}{*}{$0.35$} &$150$ & $5$ & \textbf{0.254} & \textbf{0.849} & \textbf{0.391} & \multirow{2}{*}{$219,826$} & $735,052$ \\
        && BSS & & $125$ & $6$ & $0.176$ & $0.449$ & $0.253$ & & $561,368$\\

    \cmidrule{2-11}
        &\multirow{2}{*}{\textbello{Orkut}}
        & BMH & \multirow{2}{*}{$0.4$} & $150$ & $6$ & \textbf{0.239} & \textbf{0.63} & \textbf{0.346} & \multirow{2}{*}{$182,285$} & $490,682$\\
       && BSS & & $150$ & $6$ & $0.232$ & $0.303$ & $0.263$ & & $243,158$\\

    \cmidrule{2-11}
        &\multirow{2}{*}{\textbello{YouTube}}
        & BMH & \multirow{2}{*}{$0.1$} & $700$ & $3$ & $0.67$ & \textbf{0.38} & \textbf{0.485} & \multirow{2}{*}{$470,216$} & $587,936$\\
        && BSS & & $420$ & $5$ & \textbf{0.698} & $0.135$ & $0.227$ & & $91,358$ \\
    \bottomrule

    \end{tabular}
    \caption{Results of ACP using BMH and BSS sketches for the datasets of \Cref{tab:ball_1,tab:ball_2} in \Cref{apx:datasets_description}. In bold are reported the best values of Precision, Recall and F1 for each dataset.
    }
    \label{tab:APC_results}
\end{table*}
}{}

\section{Future Work}\label{sec:conclusion}
In \cite{pang2024similarity}, the authors conclude by explicitly posing the open issue ``\textit{how to dynamically
maintain the MinHash signature}'' (of the neighborhoods of a large input graph) in order to extend their algorithm for identifying
large quasi-cliques to \textit{dynamic} graphs.
An interesting future work would be to adopt our $\ell$-buffered \kminht\ solution for this problem (and other graph-mining problems) when the input graph changes over time according to a stream of update operations (\eg\ edge insertions and deletions).

A further interesting open question is whether the buffering technique can be efficiently applied  to other similarity notions such as Cosine and Hamming ones \cite{Leskobook20,prezza2024algorithmsmassivedata}.



\iffull{}{\newpage}
\bibliographystyle{plain}
\balance
\bibliography{sketch.bib}

\begin{thebibliography}{10}

\bibitem{LiveJournal1}
Lars Backstrom, Dan Huttenlocher, Jon Kleinberg, and Xiangyang Lan.
\newblock Group formation in large social networks: membership, growth, and evolution.
\newblock In {\em Proceedings of the 12th ACM SIGKDD International Conference on Knowledge Discovery and Data Mining}, KDD '06, page 44–54, New York, NY, USA, 2006. Association for Computing Machinery.

\bibitem{baeza99modern}
Ricardo~A. Baeza-Yates and Berthier~A. Ribeiro-Neto.
\newblock {\em Modern Information Retrieval}.
\newblock ACM Press / Addison-Wesley, 1999.

\bibitem{blum2020foundations}
Avrim Blum, John Hopcroft, and Ravindran Kannan.
\newblock {\em Foundations of data science}.
\newblock Cambridge University Press, 2020.

\bibitem{broder1997resemblance}
Andrei~Z Broder.
\newblock On the resemblance and containment of documents.
\newblock In {\em Proceedings. Compression and Complexity of SEQUENCES 1997 (Cat. No. 97TB100171)}, pages 21--29. IEEE, 1997.

\bibitem{broder00identifying}
Andrei~Z Broder.
\newblock Identifying and filtering near-duplicate documents.
\newblock In {\em Annual symposium on combinatorial pattern matching}, pages 1--10. Springer, 2000.

\bibitem{BCFM00}
Andrei~Z Broder, Moses Charikar, Alan~M Frieze, and Michael Mitzenmacher.
\newblock Min-wise independent permutations.
\newblock {\em Journal of Computer and System Sciences}, 60(3):630--659, 2000.

\bibitem{BSS20}
Marc Bury, Chris Schwiegelshohn, and Mara Sorella.
\newblock Similarity search for dynamic data streams.
\newblock {\em IEEE Transactions on Knowledge and Data Engineering}, 32:2241--2253, 2020.

\bibitem{Plos18}
Benjamin~Paul Chamberlain, Josh Levy-Kramer, Clive Humby, and Marc~Peter Deisenroth.
\newblock Real-time community detection in full social networks on a laptop.
\newblock {\em PLOS ONE}, 13(1):1--37, 01 2018.

\bibitem{charikar2002similarity}
Moses~S Charikar.
\newblock Similarity estimation techniques from rounding algorithms.
\newblock In {\em Proceedings of the thiry-fourth annual ACM symposium on Theory of computing}, pages 380--388, 2002.

\bibitem{CDJ+00}
E.~Cohen, M.~Datar, S.~Fujiwara, A.~Gionis, P.~Indyk, R.~Motwani, J.D. Ullman, and C.~Yang.
\newblock Finding interesting associations without support pruning.
\newblock In {\em Proceedings of 16th International Conference on Data Engineering (Cat. No.00CB37073)}, pages 489--500, 2000.

\bibitem{CK07}
Edith Cohen and Haim Kaplan.
\newblock Summarizing data using bottom-k sketches.
\newblock In {\em ACM SIGACT-SIGOPS Symposium on Principles of Distributed Computing}, 2007.

\bibitem{cormen01introduction}
Thomas~H. Cormen, Charles~E. Leiserson, Ronald~L. Rivest, and Clifford Stein.
\newblock {\em Introduction to Algorithms}.
\newblock The MIT Press, 2nd edition, 2001.

\bibitem{dahlgaard2017fast}
S{\o}ren Dahlgaard, Mathias B{\ae}k~Tejs Knudsen, and Mikkel Thorup.
\newblock Fast similarity sketching.
\newblock In {\em 2017 IEEE 58th Annual Symposium on Foundations of Computer Science (FOCS)}, pages 663--671. IEEE, 2017.

\bibitem{DynamoDB}
Mostafa Elhemali, Niall Gallagher, Nick Gordon, Joseph Idziorek, Richard Krog, Colin Lazier, Erben Mo, Akhilesh Mritunjai, Somasundaram Perianayagam, Tim Rath, Swami Sivasubramanian, James Christopher~Sorenson III, Sroaj Sosothikul, Doug Terry, and Akshat Vig.
\newblock Amazon {DynamoDB}: A scalable, predictably performant, and fully managed {NoSQL} database service.
\newblock In {\em 2022 USENIX Annual Technical Conference (USENIX ATC 22)}, pages 1037--1048, Carlsbad, CA, July 2022. USENIX Association.

\bibitem{feigenblat-apx-kminwise-soda11}
Guy Feigenblat, Ely Porat, and Ariel Shiftan.
\newblock Exponential time improvement for min-wise based algorithms.
\newblock In {\em Proceedings of the twenty-second annual ACM-SIAM symposium on Discrete Algorithms}, pages 57--66. SIAM, 2011.

\bibitem{feigenblat2017dk}
Guy Feigenblat, Ely Porat, and Ariel Shiftan.
\newblock dk-min-wise independent family of hash functions.
\newblock {\em Journal of Computer and System Sciences}, 84:171--184, 2017.

\bibitem{HarPeled2012ApproximateNN}
Sariel Har-Peled, Piotr Indyk, and Rajeev Motwani.
\newblock Approximate nearest neighbor: Towards removing the curse of dimensionality.
\newblock {\em Theory Comput.}, 8:321--350, 2012.

\bibitem{HHW97}
Joseph~M. Hellerstein, Peter~J. Haas, and Helen~J. Wang.
\newblock Online aggregation.
\newblock In {\em ACM SIGMOD Conference}, 1997.

\bibitem{indyk1998approximate}
Piotr Indyk and Rajeev Motwani.
\newblock Approximate nearest neighbors: towards removing the curse of dimensionality.
\newblock In {\em Proceedings of the thirtieth annual ACM symposium on Theory of computing}, pages 604--613, 1998.

\bibitem{VOS}
Peng Jia, Pinghui Wang, Jing Tao, and Xiaohong Guan.
\newblock A fast sketch method for mining user similarities over fully dynamic graph streams.
\newblock In {\em 2019 IEEE 35th International Conference on Data Engineering (ICDE)}, pages 1682--1685, 2019.

\bibitem{Kathare2022ACS}
Nikita Kathare, O.~Vinati Reddy, and Dr.~Vishalakshi Prabhu.
\newblock A comprehensive study of elastic search.
\newblock {\em Journal of Research in Science and Engineering}, 2022.

\bibitem{kobayashi2000information}
Mei Kobayashi and Koichi Takeda.
\newblock Information retrieval on the web.
\newblock {\em ACM computing surveys (CSUR)}, 32(2):144--173, 2000.

\bibitem{langville2006google}
Amy~N Langville and Carl~D Meyer.
\newblock {\em Google's PageRank and beyond: The science of search engine rankings}.
\newblock Princeton university press, 2006.

\bibitem{LiveJournal2}
Jure Leskovec, Kevin Lang, Anirban Dasgupta, and Michael Mahoney.
\newblock Community structure in large networks: Natural cluster sizes and the absence of large well-defined clusters.
\newblock {\em Internet Mathematics}, 6, 11 2008.

\bibitem{Leskobook20}
Jure Leskovec, Anand Rajaraman, and Jeffrey~David Ullman.
\newblock {\em Mining of massive data sets}.
\newblock Cambridge university press, 2020.

\bibitem{LOZ12}
Ping Li, Art Owen, and Cun-hui Zhang.
\newblock One permutation hashing.
\newblock In F.~Pereira, C.J. Burges, L.~Bottou, and K.Q. Weinberger, editors, {\em Advances in Neural Information Processing Systems}, volume~25. Curran Associates, Inc., 2012.

\bibitem{manber94shing}
Udi Manber et~al.
\newblock Finding similar files in a large file system.
\newblock In {\em Usenix winter}, volume~94, pages 1--10, 1994.

\bibitem{ANF02}
Christopher~R. Palmer, Phillip~B. Gibbons, and Christos Faloutsos.
\newblock Anf: a fast and scalable tool for data mining in massive graphs.
\newblock {\em Proceedings of the eighth ACM SIGKDD international conference on Knowledge discovery and data mining}, 2002.

\bibitem{pang2024similarity}
Jiayang Pang, Chenhao Ma, and Yixiang Fang.
\newblock A similarity-based approach for efficient large quasi-clique detection.
\newblock In {\em Proceedings of the ACM on Web Conference 2024}, pages 401--409, 2024.

\bibitem{PR00}
Parag~C. Pendharkar and James~A. Rodger.
\newblock Data mining using client/server systems.
\newblock {\em Journal of Systems and Information Technology}, 4:72--82, 2000.

\bibitem{Dynamic_minwise}
Rameshwar Pratap and Raghav Kulkarni.
\newblock Minwise-independent permutations with insertion and deletion of features.
\newblock In Oscar Pedreira and Vladimir Estivill-Castro, editors, {\em Similarity Search and Applications}, pages 171--184, Cham, 2023. Springer Nature Switzerland.

\bibitem{prezza2024algorithmsmassivedata}
Nicola Prezza.
\newblock Algorithms for massive data -- lecture notes, 2024.

\bibitem{tabulation}
Mihai Pundefinedtra\c{s}cu and Mikkel Thorup.
\newblock The power of simple tabulation hashing.
\newblock {\em J. ACM}, 59(3), jun 2012.

\bibitem{Tarjan_amortized}
Robert~Endre Tarjan.
\newblock Amortized computational complexity.
\newblock {\em SIAM Journal on Algebraic Discrete Methods}, 6(2):306--318, 1985.

\bibitem{Tho-bottomk-13}
Mikkel Thorup.
\newblock Bottom-k and priority sampling, set similarity and subset sums with minimal independence.
\newblock In {\em Proceedings of the Forty-Fifth Annual ACM Symposium on Theory of Computing}, STOC '13, page 371–380, New York, NY, USA, 2013. Association for Computing Machinery.

\bibitem{MROS}
Qingjun Xiao, Shiwei Yang, Panpan Li, Kangying Li, and Lin Wen.
\newblock Multi-resolution odd sketch for mining extended jaccard similarity of dynamic streaming sets.
\newblock {\em IEEE Transactions on Network Science and Engineering}, pages 1--15, 2023.

\bibitem{YouTube-Orkut}
Jaewon Yang and Jure Leskovec.
\newblock Defining and evaluating network communities based on ground-truth.
\newblock In {\em Proceedings of the ACM SIGKDD Workshop on Mining Data Semantics}, MDS '12, New York, NY, USA, 2012. Association for Computing Machinery.

\bibitem{Zobrist1990ANH}
Albert~L. Zobrist.
\newblock A new hashing method with application for game playing.
\newblock {\em ICGA Journal}, 13:69--73, 1990.

\end{thebibliography}

\iffull{
\newpage
\appendix

\clearpage

\section{Further Related Work} \label{sec:related}

For brevity' sake, in what follows we only discuss those previous results that are strongly related to our setting.

\paragraph{Fully-Dynamic Sketches for SE Queries.}
In what follows, we provide a short overview of efficient sketch solutions for SE queries in fully-dynamic setting while they do not yield LSH families and, thus, cannot be applied  for efficient  ACP queries.

In \cite{BSS20}, besides the global solution discussed in the previous subsection, another sketch scheme is presented for SE operations that achieves   additive approximation error $\varepsilon$ with constant probability, while using $O((\frac{1}{\varepsilon^2}) m \log N)$ memory space. The update and the query time is constant. If one wants to guarantee this error bound  for \textit{all} pairs of sets and/or improve the success probability, then it is necessary to maintain independent copies of this sketch at a cost of an additional multiplicative factor of $O(\log m)$ in the space usage, query and update time.

In \cite{VOS}, a data structure VOS is proposed that maintains subset sketches over \textit{legal} streams of element insertions and deletions.  
This virtual sketch takes $O(1)$ update and $O(k)$ query time, where $k$ is a parameter that controls the quality of the approximation. The sketch can answer SE queries, namely can approximate the Jaccard similarity of any two subsets via an estimation of the size of their symmetric difference. The expected additive error and its variance are bounded by quite involved formulas expressed in terms of $k$ and the total used space.

In \cite{MROS}, a better version of VOS, called \textit{Multi-Resolution Odd Sketch} (MROS), is obtained by reducing the memory space while slightly improving the quality of the Jaccard similarity estimation. The update time is equivalent to that of VOS.

\paragraph{Similarity Sketches for Other Dynamic Input Models.}
In \cite{feigenblat-apx-kminwise-soda11}, by constructing small approximate $k$-min-wise independent functions (see \cref{def:apx_min-hash},) a sketch scheme is derived which is able to support efficient SE queries over two types of input streaming models: \textit{windowed} and \textit{incremental}. In the latter, only element insertions are allowed while, in the former, the data sketch refers only to the data determined by the last $\Delta$ insertions. For these input models, their sketch scheme yields to an exponential time improvement \wrt\ previous solutions. In particular, the used memory space (\ie\ number of words) is $O(k \log\log (1/\varepsilon) + \log(1/\varepsilon) )$ and the amortized hashing time is $O(\log^{2}(k \log\log (1/\varepsilon) + \log(1/\varepsilon)))$, while the additional time per item is $O(\log k)$.
Significant improvements of the approach in \cite{feigenblat-apx-kminwise-soda11}, in terms of memory space, have been recently obtained in \cite{feigenblat2017dk}.

\paragraph{Alternative Similarity Sketches to  Min-Hashing.}
In the literature, several alternative approaches to Min-Hashing have been proposed, either improving the running time or the space required. Here we just briefly mention the most popular ones.
In \cite{LOZ12}, the authors introduced the \textit{One-Permutation Hashing} (OPH) scheme:
this sketch partition the elements of a set $A$ into $k$ buckets using a single hash function, then the signature is the minimum hash value of each bucket.
This sketch requires $O(k)$ memory words and can be computed in linear time $O(|A|)$, thus improving the $O(k|A|)$ time complexity required to the Min-Hashing scheme. Unfortunately, it suffers the so-called \textit{empty-bucket problem} \cite{dahlgaard2017fast} when the set has small size \wrt\ $k$. A more  efficient sketch  is that introduced in \cite{dahlgaard2017fast}.
It has similar performance of OPH but it overcomes the empty-bucket problem. The analysis of the running time strongly relies on the fact that the sets are given off-line, and it is not clear if such efficiency can be guaranteed for dynamic data streams.
Another popular and widely used similarity-preserving sketch is the \textit{bottom-$k$ sketch} \cite{CK07, HHW97, Tho-bottomk-13}. This uses $O(k)$ memory words, can be computed in $O(|A| \log{k})$, and uses a single hash function to select the $k$ minimum hash values.
This sketch can be  used to estimate the Jaccard similarity of any two sets with an expected additive error of $O(1/\sqrt k)$, although the result does not hold in concentration and, most importantly, it cannot be used to get efficient ACP \cite{dahlgaard2017fast}.

\section{Experimental Details}\label{apx:other_experiments}

\subsection{Implementation Details}\label{apx:implementation}
Since the state-of-the-art solutions require a \emph{legal} stream, we implemented a simplified version of the BMH sketch that works for \emph{legal} streams only.

The BMH sketch is implemented as an array of $k$ $\ell$-buffers, where each buffer stores the hash values only.
Since in our experiments the value of $\ell$ is reasonably small, instead of using a balanced binary search tree, we implement the $\ell$-buffers as unordered arrays. To speed-up the operation \textbello{get-signature}, we explicitly keep the current \kminht\ signature updated after each operation. Notice that these changes would not affect the theoretical asymptotic bounds stated in \Cref{thm:main_result}.

We implemented both versions\footnote{No actual implementation is available.} of the sketch in \cite{BSS20}: BSS and BSS-proactive sketches.
The BSS sketch consists of  a matrix of $\log{|U|}$ rows (say $T^{A}_i$) each of $c^2$ integers, two $2$-wise independent hash functions $h_1, h_2$ and the current size $n_A = \vert A \vert$ of the set.
The update operations are performed in constant time using the two hash functions $h_1,h_2$ to update the right row and updating the parameter $n_A$ (see Algorithm $1$ of \cite{BSS20}).
To answer the query we first consider the $(\log{(\alpha \cdot n_A)})$-th row as a characteristic vector and then we compute the \kminht\ signature in $O(kc^2)$ time.
We fix  $\alpha = 0.1$ according to  the setting in \cite{BSS20, MROS}.

As we already mentioned in the introduction, to speed up the query time, \cite{BSS20} proposed a second version of the sketch called BSS-proactive.
The BSS-proactive dynamically maintains the \kminht\ signatures of each $T^{A}_i$ and updates them after each update operation as in our BMH sketch, instead of computing it from scratch at query time.
Notice that the BSS-proactive suffers of the same problem of \minht, \ie\ a deletion may cause a fault affecting the \kminht\ signature of some $T^{A}_i$. However, the performance of BSS-proactive is better than the \vanilla\ one, since the re-computation of the signature of $T^{A}_i$ is executed on a single row of size $c^2$ instead of the whole set $A$: the BSS-proactive worst-case time needed to perform a deletion is in fact $O(k c^2)$.



For the sake of a fair comparison, we use the same hash functions to compute the signatures of the sets (in our solution) and the signatures of the $T^A_i$s (for BSS sketch). We re-sample new hash functions at each experiment.
We use \textit{Tabulation Hashing} \cite{tabulation,Zobrist1990ANH} with $8$ tables each of $16$ entries.
We point out that Tabulation Hashing may be evaluated in $O(1)$ time and it is (only) $3$-wise independent. 
Even though our analysis of the $\ell$-buffered \kminht\ requires a stronger hash family, in practice the experiments show that Tabulation Hashing is enough to achieve good performance and a small number of faults.

As for pseudo-random generators, we adopt the Marsenne Twister provided by the standard \textbello{c++} library\footnote{\url{https://en.cppreference.com/w/cpp/numeric/random/mersenne_twister_engine}}.

Our implementation is publicly available at \url{https://github.com/Alessandrostr95/FullyDynamicKMinHash}.

\subsection{Datasets}\label{apx:datasets_description}
We use two types of synthetic datasets. In the first one, that we use to measure the execution time of the considered solutions, we simply sample uniformly at random subsets of different sizes from the universe $U = \left[ 2^{32} \right]$, and store them with a sparse representation.

To compare the precision of the Jaccard similarity estimation obtained by the different sketches,  we generate a different type of  synthetic dataset by adopting the benchmark of \cite{CDJ+00}. In more detail, a set $A \subseteq U$ is represented as a characteristic vector, \ie\ a bit vector of size $|U|$ where a $1$ in position $j$ indicates that $A$ includes the $j$-th item.
We then generate a set by setting each entry to $1$ with probability $q = 0.05$.
In this way we obtain on average sets of size $5\%$ of $|U|$.
For each set $A$ we further generate a second set $A'$ by flipping each $1$ of $A$ with probability $p_1$ and each $0$ with probability $p_2$ as in \cite{CDJ+00}.
We choose $p_1,p_2$ such that the size of $A'$ is almost the same size of $A$ and the Jaccard similarity $\Jsim(A,A')$ is, in expectation, a given parameter $J$.

As in \cite{VOS,pang2024similarity, Plos18}, we evaluate ACP queries among neighborhoods of nodes in real networks such as LiveJournal~\cite{LiveJournal1,LiveJournal2}, Youtube~\cite{YouTube-Orkut} and Orkut~\cite{YouTube-Orkut}.
For each network, we select the $5000$ vertices with the highest out-degree, and, for each such a vertex, we compute the \emph{ball} of radius $d$ centered in the vertex, \ie\ all the vertices at hop distance at most $d$. In particular, we consider the cases $d=1$ and $d=2$.  \Cref{tab:dataset_statistics,tab:ball_1,tab:ball_2} provide a summary of the generated datasets, and for each dataset, information regarding the number of pairs with Jaccard similarity above a threshold $J$.

\begin{table}[ht]
    \centering
    \begin{tabular}{ll|ccc}
    \toprule
         & Statistics & \textbello{LiveJournal} & \textbello{Orkut} & \textbello{YouTube} \\
    \midrule
        & Nodes & 4,847,571 & 3,072,441 & 1,134,890 \\
        & Edges & 68,993,773 & 117,185,083 & 2,987,624 \\
        \midrule

        \multirow{3}{*}{$d=1$} & Max size & 20,293 & 33,007 & 28,576 \\

        & Min size & 396 & 753 & 75 \\

        & Average size & 570 & 1,719 & 217 \\
        \midrule

       \multirow{3}{*}{$d=2$} & Max size & 318,359 & 936,275 & 333,261 \\

        & Min size & 489 & 6,507 & 82 \\

        & Average size & 32,903 & 153,151 & 8,998
        \\
    \bottomrule
    \end{tabular}
    \caption{For each network, the number of nodes, the number of edges, and the maximum size, minimum size and average size of the considered balls of radius $d=1$ and $d=2$.}
    \label{tab:dataset_statistics}
\end{table}

\begin{table}[ht]
    \centering
    \begin{tabular}{c|ccc}
    \toprule
         $J$ & \textbello{LiveJournal} & \textbello{Orkut} & \textbello{YouTube} \\
    \midrule
        $0.0$ & \makecell{12,497,500 \\ $100\%$} & \makecell{12,497,500 \\ $100\%$} & \makecell{12,497,500 \\ $100\%$} \\ \hline

        $0.05$ & \makecell{188,847 \\ $1.51\%$} & \makecell{102,841 \\ $0.82\%$} & \makecell{23,763 \\ $0.19\%$} \\ \hline

        $0.1$ & \makecell{113,117 \\ $0.91\%$} & \makecell{12,604 \\ $0.1\%$} & \makecell{3,801 \\ $0.03\%$} \\ \hline

        $0.15$ & \makecell{96,199 \\ $0.77\%$} & \makecell{3,150 \\ $0.03\%$} & \makecell{901 \\ $0.007\%$} \\ \hline

        $0.2$ & \makecell{88,849 \\ $0.71\%$} & \makecell{1,231 \\ $0.01\%$} & \makecell{251 \\ $0.002\%$} \\ \hline

        $0.25$ & \makecell{84,316 \\ $0.67\%$} & \makecell{554 \\ $0.004\%$} & \makecell{82 \\ $0.0007\%$} \\ \hline

        $0.3$ & \makecell{81,046 \\ $0.65\%$} & \makecell{277 \\ $0.002\%$} & \makecell{31 \\ $0.0002\%$} \\ \hline

        $0.35$ & \makecell{78,020 \\ $0.62\%$} & \makecell{140 \\ $0.001\%$} & \makecell{13 \\ $0.0001\%$} \\ \hline

        $0.4$ & \makecell{75,114 \\ $0.60\%$} & \makecell{78 \\ $0.0006\%$} & \makecell{9 \\ $0.0001\%$} \\ \hline

        $0.45$ & \makecell{71,526 \\ $0.57\%$} & \makecell{42 \\ $0.0003\%$} & \makecell{5 \\ $0.0\%$} \\ \hline

        $0.50$ & \makecell{67,599 \\ $0.54\%$} & \makecell{20 \\ $0.0002\%$} & \makecell{4 \\ $0.0\%$} \\
    \bottomrule
    \end{tabular}
    \caption{For each network, we select the $5000$ vertices with the highest out-degree and for each of them we compute the set of vertices at hop distance at most $1$.
    For different values of similarity $J$, the number of pairs whose similarity is $\geq J$ is then reported.}
    \label{tab:ball_1}
\end{table}

\begin{table}[ht]
    \centering
    \begin{tabular}{c|ccc}
    \toprule
         $J$ & \textbello{LiveJournal} & \textbello{Orkut} & \textbello{YouTube} \\
    \midrule
        $0.0$ & \makecell{12,497,500 \\ $100\%$} & \makecell{12,497,500 \\ $100\%$} & \makecell{12,497,500 \\ $100\%$} \\ \hline

        $0.05$ & \makecell{4,182,176 \\ $33.46\%$} & \makecell{8,812,313 \\ $70.51\%$} & \makecell{1,122,642 \\ $8.98\%$} \\ \hline

        $0.1$ & \makecell{3,851,796 \\ $30.82\%$} & \makecell{6,957,490 \\ $55.67\%$} & \makecell{470,216 \\ $3.76\%$} \\ \hline

        $0.15$ & \makecell{3,481,505 \\ $27.86\%$} & \makecell{5,301,892 \\ $42.42\%$} & \makecell{223,122 \\ $1.79\%$} \\ \hline

        $0.2$ & \makecell{2,728,234 \\ $21.83\%$} & \makecell{3,812,368 \\ $30.50\%$} & \makecell{11,6701 \\ $0.94\%$} \\ \hline

        $0.25$ & \makecell{1,511,979 \\ $12.10\%$} & \makecell{245,0317 \\ $19.61\%$} & \makecell{68,421 \\ $0.55\%$} \\ \hline

	$0.3$ & \makecell{604,608 \\ $4.84\%$} & \makecell{1,304,592 \\ $10.44\%$} & \makecell{43,219 \\ $0.35\%$} \\ \hline

	$0.35$ & \makecell{219,826 \\ $1.76\%$} & \makecell{555,262 \\ $4.44\%$} & \makecell{27,997 \\ $0.22\%$} \\ \hline

	$0.4$ & \makecell{116,706 \\ $0.93\%$} & \makecell{182,285 \\ $1.45\%$} & \makecell{17,374 \\ $0.14\%$} \\ \hline

	$0.45$ & \makecell{95,208 \\ $0.76\%$} & \makecell{52,261 \\ $0.42\%$} & \makecell{10,235 \\ $0.08\%$} \\ \hline

	$0.5$ & \makecell{88,422 \\ $0.71\%$} & \makecell{14,146 \\ $0.11\%$} & \makecell{5,576 \\ $0.04\%$} \\
    \bottomrule
    \end{tabular}
    \caption{For each network, we select the $5000$ vertices with the highest out-degree and for each of them we compute the set of vertices at hop distance at most $2$.
    For different values of similarity $J$, the number of pairs whose similarity is $\geq J$ is then reported.}
    \label{tab:ball_2}
\end{table}

\begin{figure*}[ht]
\begin{subfigure}{.3\textwidth}
  \centering
  \includegraphics[width=.97\linewidth]{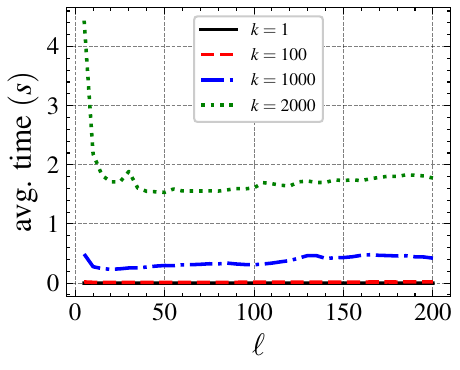}
  \caption{$n = 2^{12}$}
  \label{fig:test_ell_small_appendix}
\end{subfigure}
\begin{subfigure}{.3\textwidth}
  \centering
  \includegraphics[width=\linewidth]{images/test_L_medium-array.pdf}
  \caption{$n = 2^{16}$}
  \label{fig:test_ell_medium_appendix}
\end{subfigure}
\begin{subfigure}{.3\textwidth}
  \centering
  \includegraphics[width=1.03\linewidth]{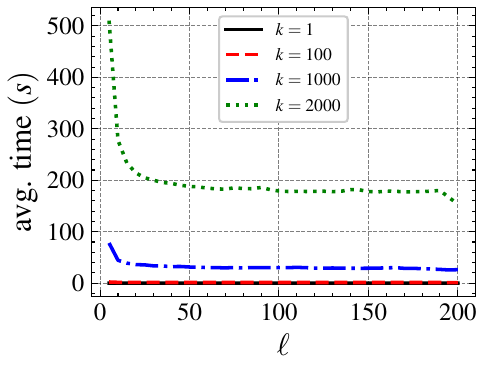}
  \caption{$n = 2^{19}$}
  \label{fig:test_ell_large_appendix}
\end{subfigure}

\begin{subfigure}{.3\textwidth}
  \centering
  \includegraphics[width=\linewidth]{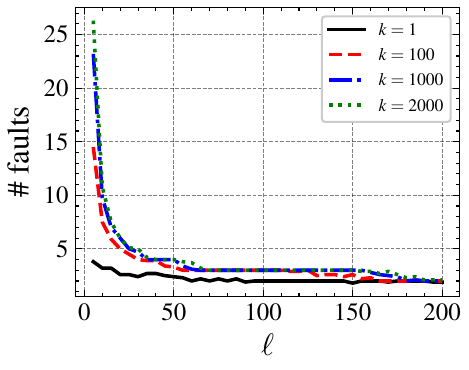}
  \caption{$n = 2^{12}$}
  \label{fig:test_ell_small_faults_appendix}
\end{subfigure}%
\begin{subfigure}{.3\textwidth}
  \centering
  \includegraphics[width=\linewidth]{images/test_L_medium_fault.pdf}
  \caption{$n = 2^{16}$}
  \label{fig:test_ell_medium_faults_appendix}
\end{subfigure}
\begin{subfigure}{.3\textwidth}
  \centering
  \includegraphics[width=\linewidth]{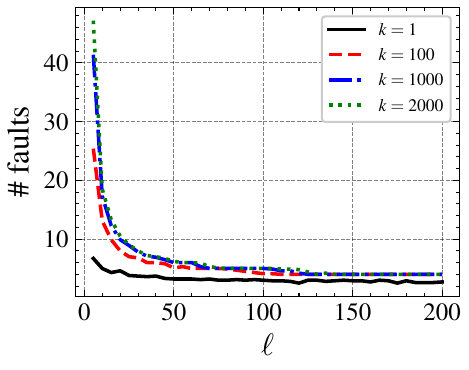}
  \caption{$n = 2^{19}$}
  \label{fig:test_ell_large_faults_appendix}
\end{subfigure}
\caption{(a), (b) and (c)
report the average execution time of $100$ independent experiments consisting in $n$ insertion followed by $n$ deletion, for values of $k = 1, 100, 1000, 2000$ and varying $\ell$ from $2$ to $200$.
Similarly, 
(d), (e) and (f) report the average number of faults for the same experiments.}
\Description{...}
\label{fig:time_ell_appendix}
\end{figure*}

\begin{figure*}[ht]
\begin{subfigure}{.3\textwidth}
  \centering
  \includegraphics[width=\linewidth]{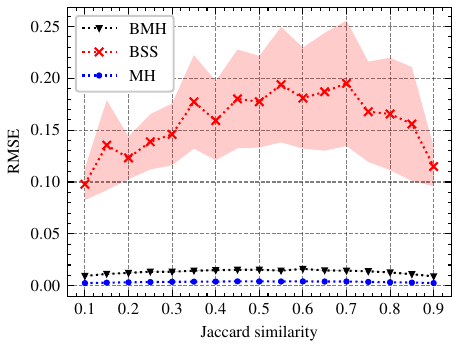}
  \caption{$|U| = 2^{14}$}
  \label{fig:exp_es_small_appendix}
\end{subfigure}
\begin{subfigure}{.3\textwidth}
  \centering
  \includegraphics[width=\linewidth]{images/se_qualityM.pdf}
  \caption{$|U| = 2^{17}$}
  \label{fig:exp_es_medium_appendix}
\end{subfigure}
\begin{subfigure}{.3\textwidth}
  \centering
  \includegraphics[width=\linewidth]{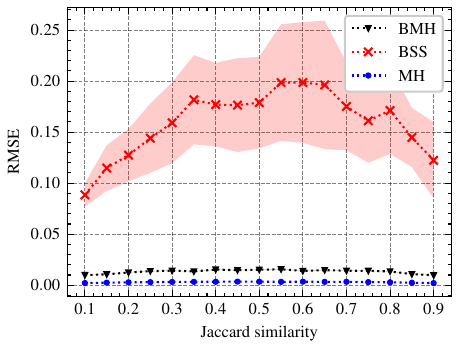}
  \caption{$|U| = 2^{20}$}
  \label{fig:exp_es_large_appendix}
\end{subfigure}
\caption{RMSE for different values of Jaccard similarity, from $0.1$ to $0.9$. The size of the sketches is $1024 \times \log{\vert U \vert}$ memory words, for different values of $|U|$. Thus \vanilla\ (MH in the figure) uses  $1024 \times \log{\vert U \vert}$ hash functions.
Each point is the RMSE for $1000$ experiments.
The shaded areas represent the standard deviation of the experiments.}
\Description{}
\label{fig:exp_se_appendix}
\end{figure*}

\subsection{Running Time Evaluation} \label{apx:running_time_evaulation}
We extend the discussion of \Cref{subsection:params} including more experiments.
The first set of experiments is devoted to empirically evaluate the time bounds proved in \Cref{sec:ourbufftech} of our data structure over streams of operations. To do so we measure the execution time needed to process sequences of operations with a special attention to the impact of the buffer size $\ell$ on the performances.
Since the query time to obtain a signature is independent from $\ell$ and from the size $n$ of the set, we consider sequences of update operations only.

In particular, we execute the following stress test. We first sample a set of $n$ elements from $U$ uniformly at random. Then, starting from $A=\emptyset$, we execute a sequence of $n$ insertions into $A$, one for each sampled element, and then an analog sequence of $n$ deletions that empties $A$.
Notice that such a sequence of operations is in some sense a bad scenario for our data structure since we execute a large number of consecutive deletions which in turn maximizes the probability of a fault.

For a given pair of $n$ and $k$, we vary $\ell$ from $1$ to $200$ and, for each chosen value of $\ell$, we perform $100$ independent experiments and report the average execution time and the average number of faults.
\Cref{fig:time_ell} shows the outcomes of the experiments for $n = 2^{16}$ and for different $k = 1, 100, 1000, 2000$ hash functions (see \Cref{fig:time_ell_appendix} in \Cref{apx:other_experiments} for different values of $n$).

\Cref{fig:test_ell_medium_faults} shows that the average number of faults decreases exponentially in $\ell$ as predicted by our theoretical analysis. Moreover, such a behaviour is not influenced by the number $k$ of the hash functions.
Even though our theoretical analysis requires $O(\log{n})$-min wise independent hash functions in order to guarantees a low fault rate, this experiments empirically suggests that Tabulation Hashing can be enough for practical purposes.

The observed execution times are reported in \Cref{fig:test_ell_medium}.
We can see that the best running time is obtained when $\ell$ is set to a value near to $\log_2{n}$ which is in line with our analysis.
Above such a threshold of $\log_2{n}$, the total running time does not improve even if the number of faults decreases. This is due to the fact that the cost saved by reducing the number of faults does not compensate for the overhead of the single update operation which increases as $\ell$ increases.
As a result, since in real applications the size of $A$ is unknown a priori, or simply changes over time, from now on we set $\ell = 32$ (\ie\ $\ell = \log_2{|U|}$).

Finally, the case $\ell=1$ deserves a special discussion, since in this case our data structure coincides with \vanilla.
Accordingly to what has been widely documented in the literature~\cite{BSS20,VOS,MROS}, we confirm that \vanilla\ performs very poorly in a fully-dynamic environment. For the sake of comparison we report in \Cref{tab:speed_up} the execution times of \vanilla\ and of our sketch with the chosen parameter $\ell = 32$ (for $k = 2000$ and $n = 2^{12}, 2^{16}, 2^{19}$). It is worth noticing that our solution strongly outperforms \vanilla, gaining a speed-up up to $745$x. This is a clear evidence that the buffering technique results in a huge running time improvement at a reasonable cost of small extra space.

\subsection{Further Experiments}\label{apx:quality_SE}
In \Cref{fig:exp_se_appendix} we report further experiments on the quality comparison between \vanilla, BMH and BSS sketches for different universe sizes $|U|$ extending \Cref{paragraph:quality_comparison}.
The experiments clearly show that the results are independent from the universe size $|U|$.

}{}
\end{document}